\documentclass[usenatbib]{mnras}
\usepackage{txfonts, color,hyperref,graphicx,amssymb,latexsym,float}
\usepackage{pdflscape}

\newcommand{\solarmass}{\,{M}_\odot}

\begin{document}

\title[Stellar spin-down and Zeeman-Doppler magnetograms]{Studying stellar spin-down with Zeeman-Doppler magnetograms}

\author[V. See et al.]
{V. See$^{1,2}$\thanks{E-mail: w.see@exeter.ac.uk}, M. Jardine$^1$, A. A. Vidotto$^3$,  J.-F. Donati$^{4,5}$, S. Boro Saikia$^{6}$,
\newauthor R. Fares$^{7}$, C. P. Folsom$^{4,5,8,9}$, \'{E}. M. H\'{e}brard$^{10}$, S. V. Jeffers$^{6}$,
\newauthor S. C. Marsden$^{11}$, J. Morin$^{12}$, P. Petit$^{4,5}$, I. A. Waite$^{11}$ and the BCool Collaboration\\
$^{1}$SUPA, School of Physics and Astronomy, University of St Andrews, North Haugh, KY16 9SS, St Andrews, UK\\
$^{2}$Department of Physics and Astronomy, University of Exeter, Physics Building, Stocker Road, Exeter, EX4 4QL, UK\\
$^{3}$School of Physics, Trinity College Dublin, University of Dublin, Dublin-2, Ireland\\
$^{4}$Universit\'{e} de Toulouse, UPS-OMP, Institut de Recherche en Astrophysique et Plan\'{e}tologie, Toulouse, France\\
$^{5}$CNRS, Institut de Recherche en Astrophysique et Plan\'{e}tologie, 14 Avenue Edouard Belin, F-31400 Toulouse, France\\
$^{6}$Universit\"at G\"ottingen, Institut f\"ur Astrophysik, Friedrich-Hund-Platz 1, 37077 G\"ottingen, Germany\\
$^{7}$INAF- Osservatorio Astrofisico di Catania, Via Santa Sofia, 78 , 95123 Catania, Italy\\
$^{8}$Univ. Grenoble Alpes, IPAG, F-38000 Grenoble, France\\
$^{9}$CNRS, IPAG, F-38000 Grenoble, France \\
$^{10}$Department of Physics and Astronomy, York University, Toronto, ON M3J 1P3, Canada\\
$^{11}$Computational Engineering and Science Research Centre, University of Southern Queensland, Toowoomba, 4350, Australia\\
$^{12}$Laboratoire Univers et Particules de Montpellier, Universit\'e de Montpellier, CNRS, F-34095, France\\}

\maketitle

\begin{abstract}
Magnetic activity and rotation are known to be intimately linked for low-mass stars. Understanding rotation evolution over the stellar lifetime is therefore an important goal within stellar astrophysics. In recent years, there has been increased focus on how the complexity of the stellar magnetic field affects the rate of angular momentum-loss from a star. This is a topic that Zeeman-Doppler imaging (ZDI), a technique that is capable of reconstructing the large-scale magnetic field topology of a star, can uniquely address.

Using a potential field source surface model, we estimate the open flux, mass loss-rate and angular momentum-loss rates for a sample of 66 stars that have been mapped with ZDI. We show that the open flux of a star is predominantly determined by the dipolar component of its magnetic field for our choice of source surface radius. We also show that, on the main sequence, the open flux, mass- and angular momentum-loss rates increase with decreasing Rossby number. The exception to this rule is stars less massive than $0.3\solarmass$. Previous work suggests that low mass M dwarfs may possess either strong, ordered and dipolar fields or weak and complex fields. This range of field strengths results in a large spread of angular momentum-loss rates for these stars and has important consequences for their spin down behaviour. Additionally, our models do not predict a transition in the mass-loss rates at the so called wind dividing line noted from Ly$\alpha$ studies.
\end{abstract}


\section{Introduction}
\label{sec:Intro}
Studies of open clusters at different ages show that the rotation periods of low mass stars ($0.1 \solarmass \lesssim M_{\star}\lesssim 1.4 \solarmass$) evolve coherently as a function of mass and age \citep{Barnes2003,Irwin2009,Barnes2010,Meibom2011,Meibom2015,Barnes2016,Stauffer2016}. While no complete theory currently exists to explain how rotation periods evolve over the stellar lifetime, a number of processes have been identified as being integral to any such theory. On the pre-main sequence, angular momentum conservation causes stars to spin up as they contract. However, star-disk interactions appear to prevent stars reaching the break-up speeds expected from contraction alone \citep{Koenigl1991,Rebull2004,Matt2010,Matt2012PMS}. Along the main sequence, stars spin down as a result of stellar winds that carry away angular momentum \citep{Weber1967}. Additionally, transport processes redistribute angular momentum within the star and can lead to core-envelope decoupling \citep{MacGregor1991,Allain1998,Bouvier2008,Spada2011}, adding a further layer of complexity to the problem.

Studying the mass-loss rates of low-mass stars and their associated spin down torques is a non-trivial task because of the diffuse nature of stellar winds. For example, the Sun has a mass-loss rate of $\sim 10^{-14}M_{\odot}{\rm yr}^{-1}$ resulting in a wind number density of only $\sim 5{\rm cm}^{-3}$ in the vicinity of Earth \citep{Balikhin1993}. With the key exception of indirect mass-loss rates estimates from Ly$\alpha$ observations (see \citet{Wood2014} and references therein), the majority of work has been theoretical by necessity. 

A number of complementary approaches exist in the literature for tackling the rotation evolution problem. One approach involves using multi-dimensional magnetohydrodynamic (MHD) simulations to determine the mass-loss rates and spin-down torques of individual stars. These can incorporate realistic magnetic field geometries at the stellar surface to give improved estimates over simulations that use idealised field geometries \citep{Vidotto2012,Vidotto2014Torque,Nicholson2016,Gomez2016}. Another approach involves deriving braking laws that predict the spin down torque as a function of fundamental stellar parameters \citep[e.g.][]{Matt2012,Cohen2014,Reville2015}. These braking laws can then be incorporated into rotation evolution models with the aim of reproducing the rotation period distributions observed in open clusters at different ages \citep{Reiners2012,Gallet2013,vanSaders2013,Brown2014,Gallet2015,Matt2015,Johnstone2015,vanSaders2016,Amard2016,Blackman2016}.

Studies have shown that a key parameter for determining the stellar spin-down torque is the open flux \citep{Mestel1987,Vidotto2014Torque,Reville2015,Reville2015b}, i.e. the flux contained in wind bearing field lines that extend away from the star. However, rotation evolution models typically only incorporate the surface magnetic field strength into their braking laws. This is partly driven by the lack of systematic studies of how open flux varies with fundamental stellar parameters. By doing so, these models neglect the topology of the magnetic fields and their effects on the rate of mass-loss and angular momentum-loss \citep{Garraffo2015}. 

One way of estimating the open flux of the star is by using a field extrapolation model \citep[e.g.][]{Jardine2002} in conjunction with a magnetogram obtained via Zeeman-Doppler imaging (ZDI). ZDI is a tomographic imaging technique that can reconstruct the large-scale component of stellar magnetic fields at the stellar surface \citep{Semel1989,Brown1991,Donati1997,Donati2006}. Previous work has already shown that fundamental stellar parameters such as internal structure \citep{Donati2008,Morin2008,Donati2009,Morin2010,Gregory2012}, rotation period \citep{Petit2008,See2015Toroidal,See2016} and age \citep{Vidotto2014Trends,Folsom2016,Rosen2016} can affect the surface magnetic field topologies of cool stars. We will build on these previous works and investigate how parameters such as open flux, mass-loss rates and angular momentum-loss rates vary with fundamental stellar parameters such as mass or rotation. 

In section \ref{sec:Model}, we outline the characteristics of the sample used in this study and the wind model we employ. In section \ref{sec:Results}, we discuss how the open flux, mass loss-rate and angular momentum-loss rate vary across our sample. Additionally, we also compare our results to those obtained from 3D MHD simulations. Concluding remarks follow in section \ref{sec:Conlusions}.

\section{Sample and wind model}
\label{sec:Model}
\begin{figure}
	\begin{center}
	\includegraphics[width=\columnwidth]{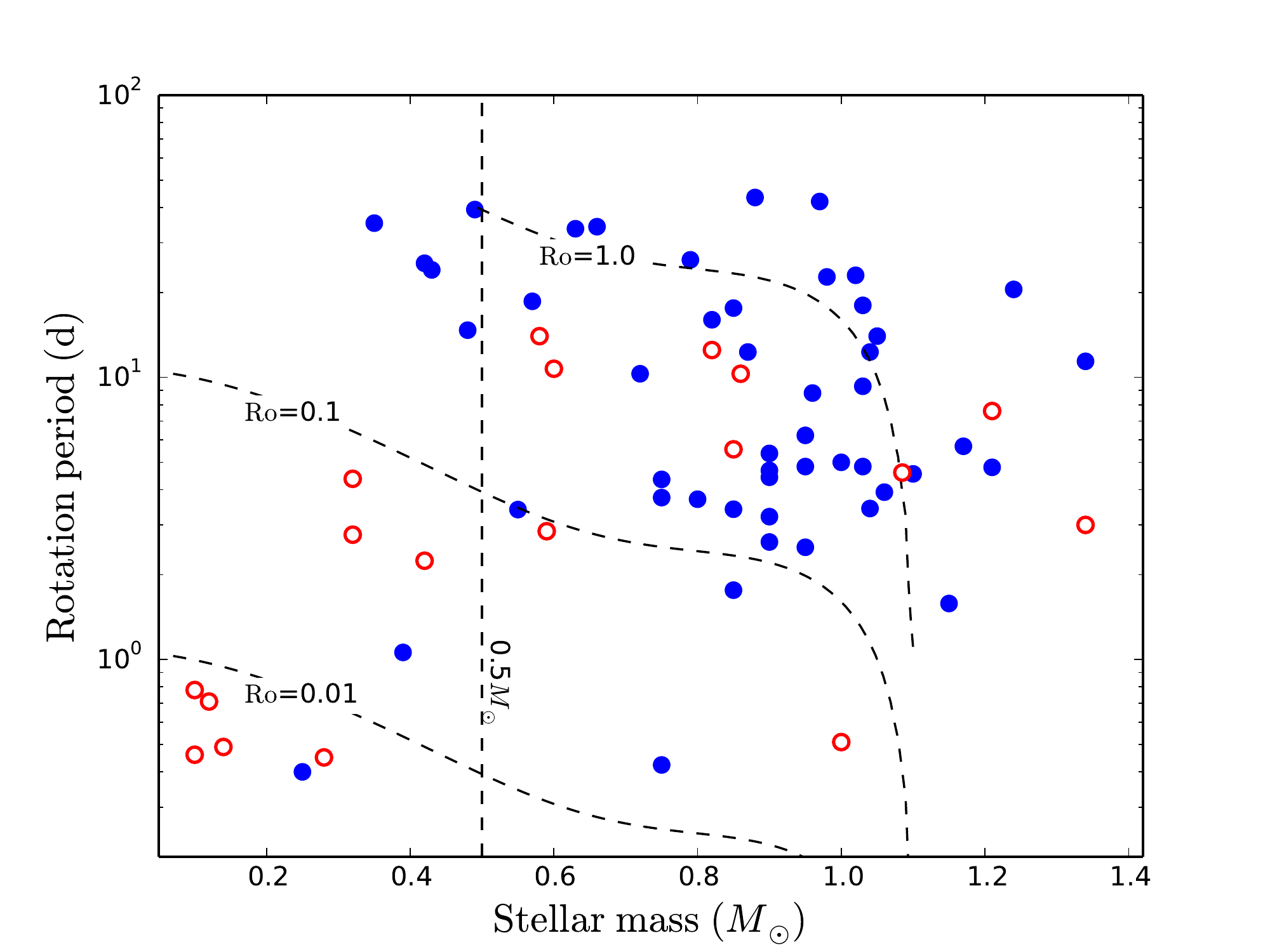}
	\end{center}
	\caption{The sample used in this work plotted in rotation period-mass parameter space. Stars with multiple ZDI maps are plotted with open red points. Dashed lines indicate a stellar mass of 0.5$M_{\odot}$ and Rossby numbers of 1.0, 0.1 and 0.01. This is an updated version of Fig. 1 from \citet{See2015Toroidal}.}
	\label{fig:ParamSpace}
\end{figure}

In this work, we use a sample of 66 stars that have each been mapped with ZDI. Many of these stars have been mapped over multiple epochs resulting in a total of 106 ZDI maps used in this work. The sample of stars used in this study is mostly comprised of the sample used by \citet{See2015Toroidal}\footnote{Our sample includes a number of stars initially presented by \citet{Folsom2016}. These stars were also included in the samples used by \citet{Vidotto2014Trends} and \citet{See2015Toroidal}. However, the masses and radii used in those works were preliminary values. We have used the updated values for these stars in this work.}. However, the sample has been expanded to include more stars presented by \citet{Folsom2016}, \citet{Hebrard2016} and H\'{e}brard et al (in prep). To date, this is the largest sample of ZDI maps used in a single study and represents well over a decade of effort observing and reconstructing the surface magnetic fields of main sequence cool dwarfs. The stars within the sample were observed under numerous programs including a large fraction from the BCool \citep{Marsden2014} and Toupies \citep{Folsom2016} collaborations. In table \ref{tab:Sample}, we list the physical parameters of each star used in this study. Stellar masses, radii and rotation periods are taken from \citet{Vidotto2014Trends} or the paper in which each ZDI map was originally published. The Rossby number is given by the rotation period divided by the convective turnover time, ${\rm Ro}=P_{\rm rot}/\tau_{\rm c}$. In this work, we will use the empirical prescription of \citet{Wright2011} (their equation (11)) to estimate the convective turnover times for our sample. The reference for the original publication of each ZDI map is also listed in table \ref{tab:Sample}. The masses and rotation periods of the sample are shown in Fig. \ref{fig:ParamSpace} (this is an updated version of Fig. 1 from \citet{See2015Toroidal}).

\subsection{Magnetic field extrapolation}
\label{subsec:Extrapolation}
Using a potential field source surface (PFSS) approach \citep{Altschuler1969}, the 3D magnetic field structure of a star can be determined. This technique has been used in numerous previous works to investigate the structure of stellar coronae \citep[e.g.][]{Jardine2002,Gregory2006,Lang2012,Johnstone2014}. By assuming that the magnetic field is in a potential state, $\underline{\nabla} \times \underline{B} = 0$, it can be defined in terms of a scalar potential, $\underline{B}=-\underline{\nabla} \psi$. Substituting into Gauss' law for magnetism, $\underline{\nabla} \cdot \underline{B} = 0$, we find that the scalar potential is the solution to Laplace's equation, $\nabla^2 \psi = 0$. The three components of the magnetic field can therefore be expressed as a sum over spherical harmonics in terms of the associated Legendre polynomials, $P_{lm}$, as follows,

\begin{equation}
	B_r=-\sum\limits_{l=1}^N \sum\limits_{m=1}^l [la_{lm} r^{l-1} - \left(l+1\right) b_{lm} r^{-\left(l+2\right)}] P_{lm} \left(\cos \theta \right) e^{im\phi}
	\label{eq:Br}
\end{equation}

\begin{equation}
	B_{\theta}=-\sum\limits_{l=1}^N \sum\limits_{m=1}^l [a_{lm}r^{l-1} + b_{lm} r^{-\left(l+2\right)}]\frac{d}{d\theta} P_{lm} \left(\cos \theta \right) e^{im\phi}
	\label{eq:Btheta}
\end{equation}

\begin{equation}
	B_{\phi}=-\sum\limits_{l=1}^N \sum\limits_{m=1}^l [a_{lm}r^{l-1} + b_{lm} r^{-\left(l+2\right)}] P_{lm} \left(\cos \theta\right) \frac{im}{\sin \theta} e^{im\phi}.
	\label{eq:Bphi}
\end{equation}
Here, $l$ indicates the spherical harmonic degree and $m$ indicates the order or `azimuthal number'. $a_{lm}$ and $b_{lm}$ are the amplitudes of each spherical harmonic component. In order to determine the values of $a_{lm}$ and $b_{lm}$, two boundary conditions are imposed; one at the stellar surface, $r_{\star}$, and one at the source surface, $r_{\rm ss}$. The stellar surface boundary is set using a ZDI map. At the source surface, the magnetic field is forced to be purely radial, i.e. $B_{\theta} = B_{\phi} = 0$. Physically, the source surface is the location beyond which all field lines are open and carrying a wind. It therefore represents the limit of coronal confinement. In this work, we set the source surface radii to be $r_{\rm ss}=3.41r_{\star}$ for the entire sample in line with previous studies \citep{Fares2010,Fares2012,See2015Radio}. It is important to note that this is a simplifying assumption and that one should expect the source surface radii to vary with the fundamental parameters of a star \citep{Reville2015b}. Indeed, to properly calculate the source surface radius, one should account for the thermal coronal energy, the bulk kinetic energy of the wind and the magnetic field energy. MHD simulations are able to self-consistently model the interactions between these components (see section \ref{subsec:Comprison} for further comparison between our models and multi-dimensional MHD models). It is therefore possible to estimate an effective source surface radii from MHD simulations. Results indicate that the source surface varies by a factor of only a few \citep[e.g.][]{Vidotto2014Torque}. The choice of a constant source surface radii is therefore a reasonable assumption, at least for this initial work. We will discuss the qualitative effects of varying the source surface radii in section \ref{subsubsec:BreakingLaws} leaving detailed exploration of the effect of varying $r_{\rm ss}$ for future work. Equations (\ref{eq:Br})-(\ref{eq:Bphi}) apply only between $r_{\star}$ and $r_{\rm ss}$. Beyond $r_{\rm ss}$, the field remains purely radial and decays as $r^{-2}$ on any given field line. The open flux of the star can then be determined by integrating the absolute radial field strength over the source surface as follows,
\begin{equation}
	\Phi_{\rm open} = \oint_{r_{\rm ss}} | B_r(r_{\rm ss})|\ {\rm d}S.
	\label{eq:Open}
\end{equation}

\subsection{Wind speed, density \& mass-loss rate}
\label{subsec:Wind}
In this work we will use a similar wind model to that of \citet{See2015Radio}. For the Sun, the wind speed is known to be correlated with the amount of field line divergence of the magnetic field lines \citep{Levine1977,Wang1990,Wang1991}. \citet{Arge2000} quantified this relationship as  

\begin{equation}
	v_{\rm wind}\left(r_{\rm Earth}\right)=267.5+\frac{410}{f_{\rm s}^{2/5}} [\mathrm{km\ s^{-1}}],
	\label{eq:WSAVelocity}
\end{equation}
where the magnetic expansion factor is

\begin{equation}
	f_{\rm s}=\left(\frac{r_{\star}}{r_{\rm ss}}\right)^2 \frac{B(r_{\star})}{B(r_{\rm ss})}.
	\label{eq:ExpansionFactor}
\end{equation}
In equation (\ref{eq:ExpansionFactor}), $B(r_{\rm ss})$ is the magnetic field strength at a given location on the source surface and $B(r_{\star})$ is the magnetic field strength at the stellar surface along the same field line. These values are determined from the field extrapolation using equations (\ref{eq:Br}) - (\ref{eq:Bphi}). Since we are studying stellar systems, we will assume that equation (\ref{eq:WSAVelocity}) gives the wind velocity at a distance of 215$r_{\star}$ - the distance at which Earth orbits the Sun.

Similar to \citet{See2015Radio} and \citet{Jardine2008}, we use a scaled solar wind density in our model. The wind density is set to be $\rho(r=215r_{\star}) = f_{\rm mag} \cdot 1.7 \times 10^{-23} {\rm g\ cm}^{-3}$ at a distance of 215$r_{\star}$ where $f_{\rm mag}$ is the average foot point strength of open field lines normalised to the average solar field strength (1G) and $1.7 \times 10^{-23} {\rm g\ cm}^{-3}$ is the solar wind density at Earth. The scaling, $f_{\rm mag}$, accounts for the denser winds of more active stars \citep{Mestel1987}. For the stars in our sample, $f_{\rm mag}$ ranges from $\sim 1$ for the least active solar-like stars to $\sim 3\times 10^3$ for the most active M dwarfs.

With knowledge of the wind speed and density, the total mass-loss rate of a star can be calculated by integrating the mass-loss rate per unit surface area over a closed surface encompassing the star beyond the source surface. Due to the way the wind speed and density are calculated, it is convenient to integrate over a spherical surface with a radius of 215$r_{\star}$, i.e.

\begin{equation}
	{\dot{M}} = \oint_{215r_{\star}} \rho v_{\rm wind}\ {\rm d}S.
	\label{eq:MassLoss}
\end{equation}
For an arbitrary stellar wind configuration, the component of the wind normal to the shell would be required to calculate the mass-loss correctly. However, since the wind flows along field lines that are radial past the source surface, the wind vector is, by definition, normal to the integrating surface in our model.

\subsection{Angular momentum-loss rates}
In this work we will make use of the formulations of \citet[][henceforth M12]{Matt2012} and \citet[][henceforth R15]{Reville2015} in order to estimate angular-momentum loss rates for our stellar sample. These authors conducted a series of multidimensional MHD simulations exploring the dependence of the angular momentum-loss rate on various parameters of the star. From these simulations, M12 found the following expression for the angular momentum-loss rate:

\begin{equation}		
\dot{J}_{\rm M12}=K_1^2B_{\star}^{4m_{1}}\dot{M}^{1-2m_{1}}r_{\star}^{4m_{1}+2}\frac{\Omega_{\star}}{(K_2^2v_{\rm esc}^2+\Omega_{\star}^2r_{\star}^2)^{m_{1}}}.
	\label{eq:Matt12}
\end{equation}
Building on the work of M12, R15 found the following expression for the angular momentum-loss rate:
\begin{figure*}
	\begin{center}
	\includegraphics[trim = 0cm 1cm 0cm 0cm, width=0.9\textwidth]{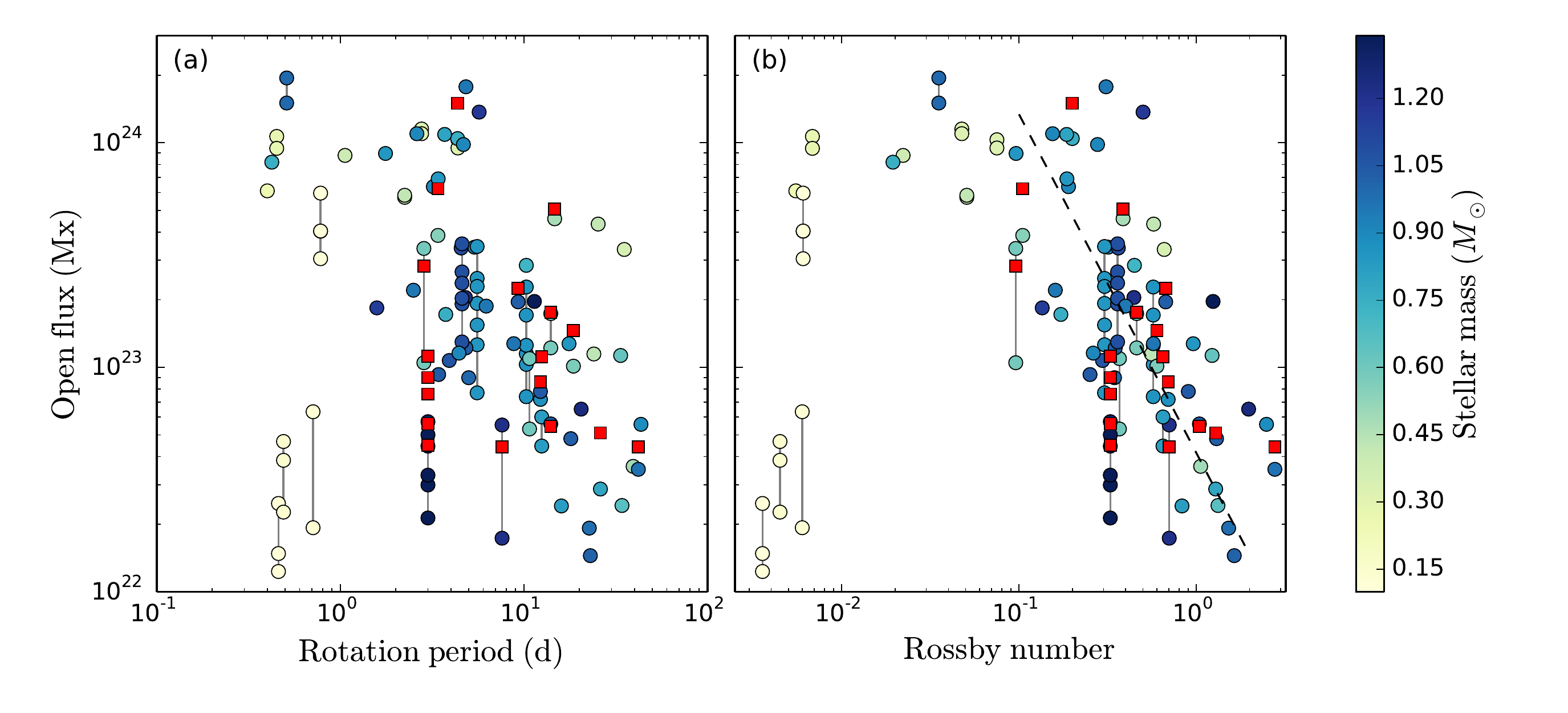}
	\end{center}
	\caption{Open flux as a function of (a) rotation period and (b) Rossby number are plotted with circle points. Stars mapped over multiple epochs are joined by a line. Each point is colour coded by stellar mass. Open fluxes calculated from self consistent 3D MHD simulations are also shown (red squares). The fit to the $\rm Ro>0.1$ stars in panel b (dashed line) has the form $\Phi_{\rm open}{\rm [Mx]} = (4.16\pm0.26)\times10^{22}\ {\rm Ro}^{-1.51\pm0.13}$. Datapoints from 3D MHD simulations (red squares) are not included in the fit. }
	\label{fig:OpenFlux}
\end{figure*}
\begin{equation}
	\dot{J}_{\rm R15}=\dot{M}\Omega_{\star}r_{\star}^2K_3^2\left(\frac{\Upsilon_{\rm open}}{\left(1+f^2/K_4^2\right)^{1/2}}\right)^{2m_2}.
	\label{eq:Reville15}
\end{equation}
In these formulations, $B_{\star}$ is the equatorial dipolar field strength at the stellar surface, $\dot{M}$ is the mass-loss rate, $r_{\star}$ is the stellar radius, $\Omega_{\star}$ is the angular rotation speed, $v_{\rm esc}=(2GM_{\star}/r_{\star})^{1/2}$ is the stellar escape velocity, $f=\Omega_{\star}r_{\star}^{3/2}(GM_{\star})^{-1/2}$ is the angular rotation speed normalised to the breakup speed, $\Upsilon_{\rm open}=\Phi^2_{\rm open}/(r_{\star}^2\dot{M}v_{\rm esc})$ is a measure of the magnetisation of the open field lines and $K_1=1.3$, $K_2=0.0506$, $K_3=0.65$, $K_4=0.06$, $m_1=0.2177$ and $m_2=0.31$ are fit parameters determined from the results of simulations\footnote{The value of $K_3$ is given as 1.4 by \citet{Reville2015}. However, this is a typographical error and the true value is $K_3=0.65$ (R\'{e}ville, priv. comm.).}. A key assumption of the M12 formulation is a dipolar magnetic field geometry while the R15 formulation encapsulates the effects of more complex field geometries. We shall explore the difference between these two formulations in section \ref{subsec:AngMomLoss}.

\section{Results}
\label{sec:Results}
\subsection{Coronal magnetic field}
\label{subsec:CoronalField}
In Fig. \ref{fig:OpenFlux}, we plot the open flux against rotation period and Rossby number, colour coded by stellar mass. The open flux values are listed in table \ref{tab:Sample}. The open flux displays a similar behaviour in both figures, increasing with decreasing rotation period/Rossby number and showing a large spread at the lowest rotation periods/Rossby numbers. However, the scatter is reduced when plotting against the Rossby number. A similar reduction in the scatter of other magnetic activity proxies is seen when they are plotted against Rossby number rather than rotation period, most notably X-ray emission \citep{Noyes1984,Pizzolato2003,Wright2011}. On the other hand, it should be noted that \citet{Reiners2014} argues that the rotation period is a more fundamental parameter than Rossby number in the context of dynamo action. 

\begin{figure*}
	\begin{center}
	\includegraphics[trim = 0cm 1cm 0cm 0cm, width=0.9\textwidth]{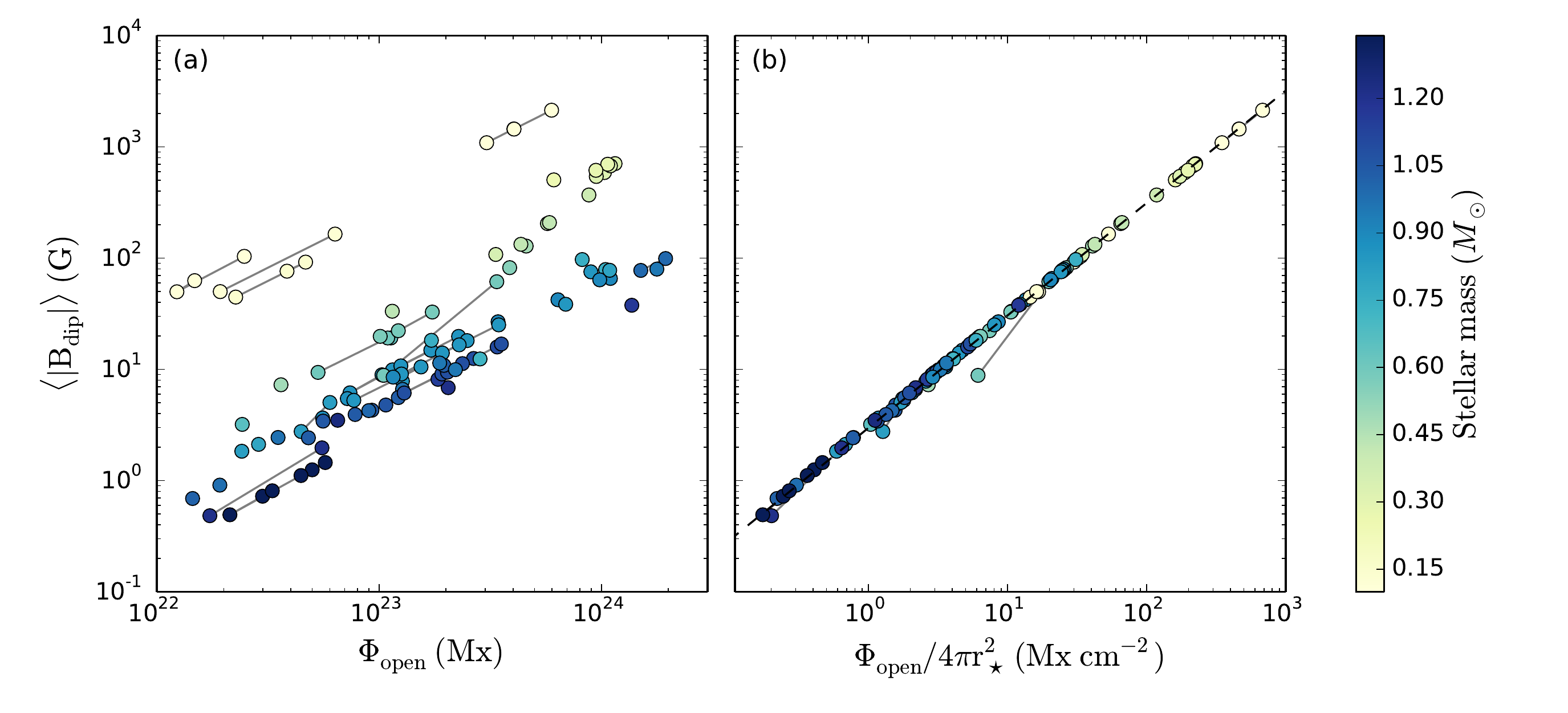}
	\end{center}
	\caption{Average unsigned magnetic field strength of the dipole component ($l=1$) at the stellar surface against (a) the open flux and (b) the open flux normalised by stellar surface area. Symbol colours and meanings are the same as Fig. \ref{fig:OpenFlux}. The dashed line in panel b is a power law fit of the form $\langle |B_{\rm dip}| \rangle = (2.97\pm0.02)(\Phi_{\rm open}/4\pi r_{\star}^2)^{(1.011\pm0.003)}$.}
	\label{fig:DipoleVsOpenFlux}
\end{figure*}

While Fig. \ref{fig:OpenFlux}b looks similar to the classical activity-rotation relationship observed in X-ray studies, there are some noticeable differences. Usually, there is a clear change in behaviour at $\rm Ro \sim 0.1$ as stars transition from the saturated to unsaturated regimes. The unsaturated regime is still evident in Fig. \ref{fig:OpenFlux}b but the saturated regime is less obvious compared to other activity proxies due to the relative dearth of stars with $\rm 0.01<Ro<0.1$ in our sample. At $\rm Ro<0.01$, a sharp change in behaviour is observed. In this regime, the open flux values show no dependence on Rossby number and are spread over two orders of magnitude. These are the lowest mass stars in our sample ($<0.3 \solarmass$). As discussed previously in the literature, low-mass M dwarfs show two distinct sets of properties; either strong dominantly dipolar fields or weak multipolar fields \citep{Morin2010}. This explains why there is such a large spread of open flux values in such a narrow Rossby number range. Given the uncertainty in the behaviour in the $\rm Ro<0.1$ regime, we perform a fit to the stars with $\rm Ro>0.1$ only. This fit has the form 

\begin{equation}
	\Phi_{\rm open}{\rm [Mx]} = (4.16\pm0.26)\times10^{22}\ {\rm Ro}^{-1.51\pm0.13}
	\label{eq:OpenFluxFit}
\end{equation}
and is shown as a dashed line in Fig \ref{fig:OpenFlux}b. All the fits in this paper used the bisector ordinary least-squares method \citep{Isobe1990}. The errors in the fit are calculated only by considering the scatter in the points without considering the intrinsic error in the individual data points. It is important to note that the form of this fit, as well as other fits presented in the rest of this work, will depend on the adopted convective turnover times. We have chosen to use the empirically determined prescription of \citet{Wright2011} but different prescriptions may produce slightly different fits. \citet{Vidotto2014Trends} determined fits for the surface flux and unsigned average magnetic field strength as a function of Rossby number for a sample of stars similar to the one used in this work. These authors tested a number of different prescriptions for the convective turnover time and found that, while the fit values changed, they did agree to within $2\sigma$ (see their appendix A5).

A number of the stars in our sample have also been modelled using self-consistent 3D MHD simulations that incorporate the same ZDI maps that we use in this work (see section \ref{subsec:Comprison} for further discussion). We have plotted the open flux values of these stars, from the MHD simulations, using red squares in Fig. \ref{fig:OpenFlux}. It is clear that the MHD values all fall within the general trend shown by the PFSS model values.

\begin{figure}
	\begin{center}
	\includegraphics[trim = 0.5cm 1.5cm 0.5cm 0cm, width=\columnwidth]{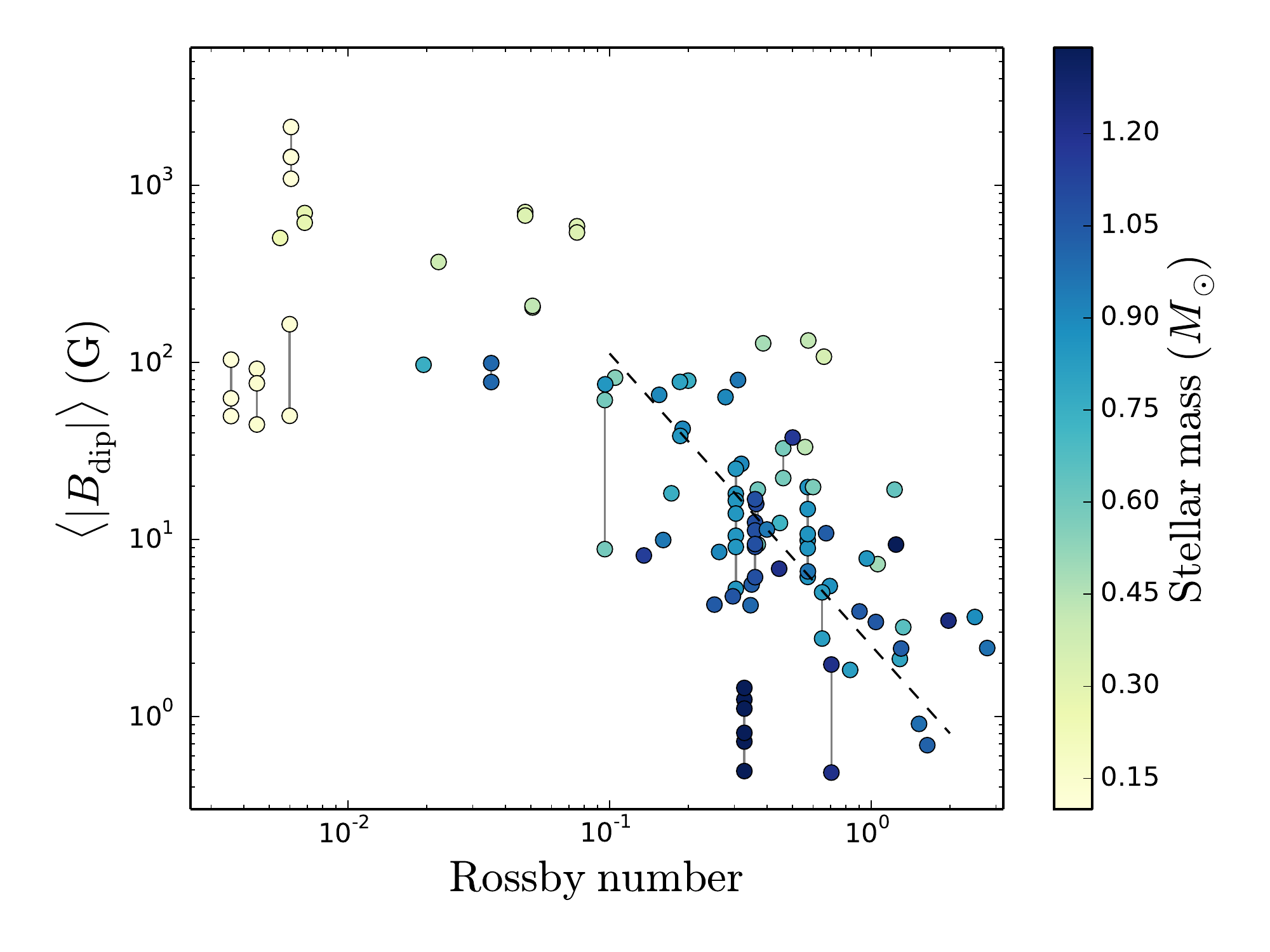}
	\end{center}
	\caption{Average unsigned magnetic field strength of the dipole component ($l=1$) at the stellar surface against Rossby number. Symbol colours and meanings are the same as Fig. \ref{fig:OpenFlux}. The fit to the $\rm Ro>0.1$ stars has the form $\langle |B_{\rm dip}|\rangle = (2.52\pm0.20)\ {\rm Ro}^{-1.65\pm 0.14}$.}
	\label{fig:DipoleRossby}
\end{figure}

It is worth considering how the resolution of the ZDI maps may affect the results of these extrapolations. The highest spherical harmonic mode that can be reconstructed by ZDI depends on the $v\sin i$ of the star in question (see \citet{Morin2010} and \citet{Fares2012} for further details). Within our sample, most maps have $l_{\rm max}$ values between 5 and 10. Jardine et al. (accepted) conducted a systematic study of the dependence of a PFSS extrapolation on the resolution of the magnetic map used. These authors took magnetograms of the Sun over two solar cycles and calculated the total magnetic flux at the stellar surface and the open flux. They repeated this process, truncating the $l_{\rm max}$ of the maps to lower values and found that the surface flux increased with higher $l_{\rm max}$ but that the open flux had very little dependence on $l_{\rm max}$. Indeed, the amount of open flux is determined predominantly by the dipole component of the Sun's magnetic field. This is not surprising given that the dipole component decays most slowly with increasing height from the stellar surface. In Fig. \ref{fig:DipoleVsOpenFlux}a, we plot the average unsigned magnetic field strength in the dipole component of the ZDI map, $\langle |B_{\rm dip}|\rangle$, against the open flux for our sample of stars. The symbol colours and meanings are the same as Fig. \ref{fig:OpenFlux}. Broadly, the trend is of increasing open flux with increasing dipole field strength. However, there is a large amount of scatter in the plot. This can be attributed to the different radii of the stars in our sample. Indeed, the influence of the stellar mass, and hence radii, can be seen in Fig. \ref{fig:DipoleVsOpenFlux}a where the colours of the points vary systematically across the plot. In Fig. \ref{fig:DipoleVsOpenFlux}b, we account for this effect by plotting the dipole field strength against the open flux normalised by the stellar surface area\footnote{The units of $\Phi_{\rm open}/4\pi r_{\star}^2$ are $\rm Mx\ cm^{-2}$. Dimensionally, this is equivalent to a magnetic field strength measured in Gauss. However, the quantity $\Phi_{\rm open}/4\pi r_{\star}^2$ does not correspond to any physically meaningful field strength and so we opt to express it in this form. Additionally, this form will be more useful for future studies since the the open flux of a star can be easily estimated from the dipole field strength and stellar radii.}. All the points collapse onto a very narrow relation given by $\langle |B_{\rm dip}| \rangle = (2.97\pm0.02)(\Phi_{\rm open}/4\pi r_{\star}^2)^{(1.011\pm0.003)}$. The fact that all the points lay on such a tight sequence and the power index of $\sim 1$ demonstrates that it is predominantly the dipole components of the magnetic field that determines the open flux of a star, at least for this choice of source surface radii. It should be expected that for sufficiently small choices of the source surface radii, higher order field modes will affect the value of the open flux (see section \ref{subsubsec:BreakingLaws} for further discussion of the impact of the source surface radius). For completeness, we plot $\langle |B_{\rm dip}|\rangle$ against Rossby number in Fig. \ref{fig:DipoleRossby}. The fit to the $\rm Ro>0.1$ stars (dashed line) has the form $\langle |B_{\rm dip}|\rangle = (2.52\pm0.20)\ {\rm Ro}^{-1.65\pm 0.14}$. Together, Figs. \ref{fig:OpenFlux}b, \ref{fig:DipoleVsOpenFlux}a and \ref{fig:DipoleRossby} show the three possible 2D projections of our sample in $(\langle |B_{\rm dip}|\rangle, \Phi_{\rm open}, {\rm Ro})$ space.

\begin{table*}
\begin{minipage}{180mm}
	\caption{Parameters for our sample: stellar mass, radius, rotation period, Rossby number, open flux, average unsigned dipolar field strength at the stellar surface, mass loss-rate, angular momentum-loss rates for the M12 and R15 formulations, instantaneous spin-down time-scale and the observation epoch. References indicate the paper where the original magnetic map was published.} 
	\label{tab:Sample}
	\begin{tabular}{lccccccccccccc}
		\hline
		Star & $M_{\star}$ & $r_{\star}$ & $P_{\rm rot}$ & $\rm Ro$	& $\Phi_{\rm open}$ & $\langle |B_{\rm dip}|\rangle$ & $\dot{M}$	& $\dot{J}_{\rm M12}$	&	$\dot{J}_{\rm R15}$	&	$\tau_{\rm R15}$ & Obs & Ref. \\
		ID & ($M_{\odot}$) & ($r_{\odot}$) & (d) & & ($10^{22}$Mx) & (G) & ($10^{-12}M_{\odot} {\rm yr}^{-1}$) & ($10^{32}$erg) & ($10^{32}$erg) & (Gyr) & epoch & \\
		\hline
\textbf{\textit{Solar-like stars}}\\																									
HD 3651	&	0.88	&	0.88	&	43.4	&	2.491	&	5.57	&	3.65	&	0.15	&	0.03	&	0.02	&	14.8	&	-	&	1	\\
HD 9986	&	1.02	&	1.04	&	23	&	1.639	&	1.45	&	0.69	&	0.04	&	0.01	&	0.01	&	125	&	-	&	1	\\
HD 10476	&	0.82	&	0.82	&	16	&	0.831	&	2.41	&	1.83	&	0.08	&	0.02	&	0.02	&	52	&	-	&	1	\\
$\kappa$ Ceti	&	1.03	&	0.95	&	9.3	&	0.673	&	19.6	&	10.9	&	0.66	&	0.86	&	0.9	&	1.94	&	2012 Oct	&	2	\\
$\epsilon$ Eri	&	0.86	&	0.77	&	10.3	&	0.572	&	17.1	&	14.9	&	0.44	&	0.45	&	0.5	&	2.76	&	2007 Jan	&	3	\\
...	&	...	&	...	&	...	&	...	&	11.5	&	9.9	&	0.34	&	0.27	&	0.28	&	5	&	2008 Jan	&	3	\\
...	&	...	&	...	&	...	&	...	&	7.4	&	6.15	&	0.32	&	0.17	&	0.16	&	8.81	&	2010 Jan	&	3	\\
...	&	...	&	...	&	...	&	...	&	10.3	&	8.93	&	0.34	&	0.25	&	0.24	&	5.73	&	2011 Oct	&	3	\\
...	&	...	&	...	&	...	&	...	&	12.5	&	10.7	&	0.4	&	0.32	&	0.33	&	4.21	&	2012 Oct	&	3	\\
...	&	...	&	...	&	...	&	...	&	22.8	&	19.8	&	0.58	&	0.67	&	0.79	&	1.74	&	2013 Oct	&	3	\\
HD 39587	&	1.03	&	1.05	&	4.8	&	0.349	&	12.2	&	5.57	&	0.77	&	1.36	&	1.09	&	3.08	&	-	&	1	\\
HD 56124	&	1.03	&	1.01	&	18	&	1.302	&	4.81	&	2.42	&	0.13	&	0.06	&	0.05	&	18.9	&	-	&	1	\\
HD 72905	&	1	&	1	&	5	&	0.346	&	8.98	&	4.26	&	0.6	&	0.78	&	0.63	&	5.02	&	-	&	1	\\
HD 73350	&	1.04	&	0.98	&	12.3	&	0.902	&	7.8	&	3.93	&	0.45	&	0.24	&	0.19	&	6.83	&	-	&	1	\\
HD 75332	&	1.21	&	1.24	&	4.8	&	0.445	&	20.5	&	6.84	&	0.61	&	2.28	&	2.12	&	1.73	&	-	&	1	\\
HD 76151	&	1.24	&	0.98	&	20.5	&	1.975	&	6.51	&	3.48	&	0.2	&	0.08	&	0.06	&	13.5	&	2007 Jan	&	4	\\
HD 78366	&	1.34	&	1.03	&	11.4	&	1.245	&	19.6	&	9.36	&	0.82	&	0.85	&	0.81	&	2.03	&	-	&	1  \\
HD 101501	&	0.85	&	0.9	&	17.6	&	0.962	&	12.7	&	7.82	&	0.37	&	0.22	&	0.23	&	3.54	&	-	&	1	\\
$\xi$ Boo A	&	0.85	&	0.84	&	5.6	&	0.304	&	34.5	&	25.1	&	1.08	&	2.81	&	3.36	&	0.75	&	-	&	1	\\
...	&	...	&	...	&	...	&	...	&	7.69	&	5.26	&	0.51	&	0.47	&	0.39	&	6.48	&	2008 Feb	&	5	\\
...	&	...	&	...	&	...	&	...	&	15.4	&	10.5	&	0.81	&	1.12	&	1.11	&	2.29	&	2009 July	&	5	\\
...	&	...	&	...	&	...	&	...	&	12.6	&	9.08	&	0.74	&	0.93	&	0.83	&	3.05	&	2010 Jan	&	5	\\
...	&	...	&	...	&	...	&	...	&	22.9	&	16.6	&	0.99	&	1.86	&	1.96	&	1.3	&	2010 Jun	&	5	\\
...	&	...	&	...	&	...	&	...	&	19.3	&	14	&	0.78	&	1.41	&	1.44	&	1.76	&	2010 Jul	&	5	\\
...	&	...	&	...	&	...	&	...	&	24.9	&	18.1	&	0.69	&	1.63	&	1.88	&	1.35	&	2011 Feb	&	5	\\
$\xi$ Boo B	&	0.72	&	1.07	&	10.3	&	0.448	&	28.4	&	12.4	&	0.98	&	1.7	&	1.89	&	0.63	&	-	&	1	\\
18 Sco	&	0.98	&	1.02	&	22.7	&	1.524	&	1.92	&	0.91	&	0.07	&	0.01	&	0.01	&	69.9	&	2007 Aug	&	4	\\
HD 166435	&	1.04	&	0.99	&	3.4	&	0.252	&	9.28	&	4.29	&	0.64	&	1.12	&	0.93	&	5.12	&	-	&	1	\\
HD 175726	&	1.06	&	1.06	&	3.9	&	0.296	&	10.7	&	4.78	&	0.55	&	1.22	&	0.99	&	4.27	&	-	&	1	\\
HD 190771	&	0.96	&	0.98	&	8.8	&	0.573	&	12.7	&	6.61	&	0.39	&	0.49	&	0.48	&	3.65	&	-	&	1	\\
61 Cyg A	&	0.66	&	0.62	&	34.2	&	1.327	&	2.42	&	3.2	&	0.06	&	0.01	&	0.01	&	59.8	&	-	&	1  \\
HN Peg	&	1.085	&	1.04	&	4.6	&	0.364	&	34	&	15.9	&	1.19	&	4.37	&	4.7	&	0.79	&	-	&	1	\\
...	&	...	&	...	&	...	&	...	&	26.6	&	12.5	&	0.82	&	2.86	&	3	&	1.21	&	2007 Jul	&	6	\\
...	&	...	&	...	&	...	&	...	&	13	&	6.13	&	0.44	&	1.08	&	0.97	&	3.74	&	2008 Aug	&	6	\\
...	&	...	&	...	&	...	&	...	&	19.1	&	9.09	&	0.6	&	1.81	&	1.77	&	2.05	&	2009 Jun	&	6	\\
...	&	...	&	...	&	...	&	...	&	23.7	&	11.3	&	0.83	&	2.63	&	2.62	&	1.39	&	2010 Jul	&	6	\\
...	&	...	&	...	&	...	&	...	&	20.3	&	9.43	&	0.8	&	2.2	&	2.12	&	1.72	&	2011 Jul	&	6	\\
...	&	...	&	...	&	...	&	...	&	35.5	&	16.9	&	1.09	&	4.36	&	4.77	&	0.76	&	2013 Jul	&	6	\\
\textbf{\textit{Young Suns}}\\																									
AB Dor	&	1	&	1	&	0.5	&	0.035	&	150	&	77.5	&	9.03	&	269	&	256	&	0.12	&	2001 Dec	&	7	\\
...	&	...	&	...	&	...	&	...	&	194	&	99.1	&	6.91	&	286	&	318	&	0.1	&	2002 Dec	&	7	\\
BD-16351	&	0.9	&	0.88	&	3.2	&	0.19	&	63.7	&	42.2	&	2.21	&	12.7	&	16	&	0.29	&	2012 Sept	&	8	\\
DX Leo	&	0.9	&	0.81	&	5.4	&	0.319	&	34.3	&	26.8	&	0.97	&	2.55	&	3.13	&	0.88	&	2014 May	&	8	\\
HII 296	&	0.9	&	0.93	&	2.6	&	0.155	&	110	&	65.6	&	3.81	&	36.1	&	48.1	&	0.12	&	2009 Oct	&	8	\\
HII 739	&	1.15	&	1.07	&	1.6	&	0.135	&	18.4	&	8.13	&	0.8	&	5.3	&	4.34	&	2.51	&	2009 Oct	&	8	\\
HIP 12545	&	0.95	&	1.07	&	4.8	&	0.31	&	178	&	79.6	&	7.19	&	52.1	&	73.3	&	0.04	&	2012 Sept	&	8	\\
HIP 76768	&	0.8	&	0.85	&	3.7	&	0.186	&	109	&	77.7	&	3.28	&	21.8	&	31.9	&	0.11	&	2013 May	&	8	\\
LO Peg	&	0.75	&	0.66	&	0.4	&	0.019	&	81.9	&	97	&	4.22	&	84.5	&	90.2	&	0.33	&	2014 Aug	&	8	\\
PELS 031	&	0.95	&	1.05	&	2.5	&	0.16	&	22	&	9.94	&	1.29	&	5.53	&	4.86	&	1.26	&	2013 Nov	&	8	\\
PW And	&	0.85	&	0.78	&	1.8	&	0.096	&	89.6	&	75.3	&	3.14	&	31.2	&	41.9	&	0.19	&	2014 Sept	&	8	\\
TYC 0486-4943-1.c	&	0.75	&	0.69	&	3.8	&	0.172	&	17.2	&	18.2	&	0.68	&	1.35	&	1.47	&	2.31	&	2013 Jun	&	8	\\
TYC 5164-567-1.c	&	0.9	&	0.89	&	4.7	&	0.277	&	98.2	&	63.8	&	2.69	&	14.7	&	21.3	&	0.15	&	2013 Jun	&	8	\\
TYC 6349-0200-1.c	&	0.85	&	0.96	&	3.4	&	0.186	&	69.1	&	38.5	&	2.15	&	14.3	&	18.2	&	0.23	&	2013 Jun	&	8	\\
TYC 6878-0195-1.c	&	1.17	&	1.37	&	5.7	&	0.501	&	137	&	37.7	&	4.78	&	37.2	&	46.5	&	0.07	&	2013 jun	&	8	\\
V439 And	&	0.95	&	0.92	&	6.2	&	0.4	&	18.7	&	11.4	&	0.62	&	1.19	&	1.21	&	2.03	&	2014 Sept	&	8	\\
V447 Lac	&	0.9	&	0.81	&	4.4	&	0.262	&	11.6	&	8.51	&	0.66	&	0.92	&	0.85	&	3.94	&	-	&	8	\\
\textbf{\textit{Hot Jupiter hosts}}\\																									
$\tau$ Boo	&	1.34	&	1.42	&	3	&	0.328	&	2.99	&	0.72	&	0.22	&	0.4	&	0.21	&	29.6	&	2008 Jan	&	9	\\
...	&	...	&	...	&	...	&	...	&	3.31	&	0.81	&	0.27	&	0.5	&	0.26	&	24.2	&	2008 Jun	&	9	\\
...	&	...	&	...	&	...	&	...	&	2.13	&	0.49	&	0.16	&	0.24	&	0.12	&	51.2	&	2008 Jul	&	9	\\
	\hline
\end{tabular}
\end{minipage}
\end{table*}

\begin{table*}
\begin{minipage}{173mm}
	\contcaption{}
	\begin{tabular}{lccccccccccccc}
		\hline
		Star & $M_{\star}$ & $r_{\star}$ & $P_{\rm rot}$ & $\rm Ro$	& $\Phi_{\rm open}$ & $\langle |B_{\rm dip}|\rangle$ & $\dot{M}$	& $\dot{J}_{\rm M12}$	&	$\dot{J}_{\rm R15}$	&	$\tau_{\rm R15}$ & Obs & Ref. \\
		ID & ($M_{\odot}$) & ($r_{\odot}$) & (d) & & ($10^{22}$Mx) & (G) & ($10^{-12}M_{\odot} {\rm yr}^{-1}$) & ($10^{32}$erg) & ($10^{32}$erg) & (Gyr) & epoch & \\
		\hline
...	&	...	&	...	&	...	&	...	&	4.45	&	1.11	&	0.24	&	0.61	&	0.35	&	17.6	&	2009 May	&	10	\\
...	&	...	&	...	&	...	&	...	&	5	&	1.25	&	0.36	&	0.86	&	0.48	&	12.9	&	2010 Jan	&	10	\\
...	&	...	&	...	&	...	&	...	&	5.72	&	1.45	&	0.25	&	0.8	&	0.5	&	12.5	&	2011 Jan	&	10 \\
HD 46375	&	0.97	&	0.86	&	42	&	2.777	&	3.51	&	2.44	&	0.09	&	0.01	&	0.01	&	36.1	&	2008 Jan	&	10	\\
HD 73256	&	1.05	&	0.89	&	14	&	1.042	&	5.59	&	3.42	&	0.22	&	0.09	&	0.08	&	15.1	&	2008 Jan	&	10	\\
HD 102195	&	0.87	&	0.82	&	12.3	&	0.695	&	7.19	&	5.46	&	0.25	&	0.14	&	0.12	&	9.47	&	2008 Jan	&	10	\\
HD 130322	&	0.79	&	0.83	&	26.1	&	1.288	&	2.87	&	2.12	&	0.08	&	0.02	&	0.01	&	40.6	&	2008 Jan	&	10	\\
HD 179949	&	1.21	&	1.19	&	7.6	&	0.704	&	1.73	&	0.48	&	0.12	&	0.05	&	0.03	&	67.7	&	2007 Jun	&	11	\\
...	&	...	&	...	&	...	&	...	&	5.53	&	1.97	&	0.24	&	0.26	&	0.19	&	12.4	&	2009 Sept	&	11	\\
HD 189733	&	0.82	&	0.76	&	12.5	&	0.649	&	4.46	&	2.76	&	0.34	&	0.07	&	0.07	&	15.5	&	2007 Jun	&	12	\\
...	&	...	&	...	&	...	&	...	&	6.01	&	5.03	&	0.32	&	0.12	&	0.1	&	10.9	&	2008 Jul	&	12  \\
\textbf{\textit{M dwarf stars}}\\																									
CE Boo	&	0.48	&	0.43	&	14.7	&	0.387	&	45.9	&	128	&	1.55	&	0.78	&	1.24	&	0.43	&	2008 Jan	&	13	\\
DS Leo	&	0.58	&	0.52	&	14	&	0.461	&	17.3	&	32.7	&	0.62	&	0.26	&	0.32	&	2.22	&	2007 Jan	&	13\\
...	&	...	&	...	&	...	&	...	&	12.2	&	22.2	&	0.44	&	0.15	&	0.18	&	3.92	&	2007 Dec	&	13	\\
GJ 182	&	0.75	&	0.82	&	4.3	&	0.2	&	104	&	78.9	&	3.59	&	18.1	&	26.6	&	0.11	&	2007 Jan	&	13	\\
GJ 49	&	0.57	&	0.51	&	18.6	&	0.6	&	10.1	&	19.8	&	0.27	&	0.07	&	0.09	&	5.88	&	2007 Jul	&	13\\
AD Leo	&	0.42	&	0.38	&	2.2	&	0.051	&	57.1	&	204	&	3.36	&	8.23	&	12.7	&	0.23	&	2007 Jan	&	14	\\
...	&	...	&	...	&	...	&	...	&	58.4	&	209	&	2.87	&	7.68	&	12.3	&	0.23	&	2008 Jan	&	14	\\
DT Vir	&	0.59	&	0.53	&	2.9	&	0.096	&	33.8	&	61.3	&	1.18	&	3.25	&	4.48	&	0.79	&	2007 Jan	&	13	\\
...	&	...	&	...	&	...	&	...	&	10.5	&	8.83	&	1.38	&	0.66	&	1.11	&	3.17	&	2007 Dec	&	13	\\
EQ Peg A	&	0.39	&	0.35	&	1.1	&	0.022	&	87.8	&	369	&	2.52	&	18.9	&	36	&	0.15	&	2006 Aug	&	14	\\
EQ Peg B	&	0.25	&	0.25	&	0.4	&	0.006	&	61.1	&	505	&	1.75	&	18.8	&	35.3	&	0.22	&	2006 Aug	&	14\\
EV Lac	&	0.32	&	0.3	&	4.4	&	0.075	&	103	&	588	&	3.45	&	5.56	&	11.8	&	0.08	&	2006 Aug	&	14	\\
...	&	...	&	...	&	...	&	...	&	94.8	&	542	&	3.04	&	4.83	&	10.2	&	0.1	&	2007 July	&	14	\\
DX Cnc	&	0.1	&	0.11	&	0.5	&	0.004	&	2.47	&	104	&	0.06	&	0.07	&	0.11	&	15.3	&	2007	&	15	\\
...	&	...	&	...	&	...	&	...	&	1.23	&	49.8	&	0.04	&	0.03	&	0.04	&	43.9	&	2008	&	15  \\
...	&	...	&	...	&	...	&	...	&	1.48	&	62.7	&	0.04	&	0.03	&	0.05	&	33.8	&	2009	&	15	\\
GJ 1156	&	0.14	&	0.16	&	0.5	&	0.004	&	2.26	&	44.6	&	0.1	&	0.12	&	0.14	&	18.6	&	2007	&	15	\\
...	&	...	&	...	&	...	&	...	&	4.67	&	92	&	0.16	&	0.28	&	0.41	&	6.44	&	2008	&	15	\\
...	&	...	&	...	&	...	&	...	&	3.85	&	76.3	&	0.13	&	0.21	&	0.3	&	8.86	&	2009	&	15	\\
GJ 1245B	&	0.12	&	0.14	&	0.7	&	0.006	&	6.34	&	164	&	0.16	&	0.23	&	0.41	&	3.52	&	2006	&	15	\\
...	&	...	&	...	&	...	&	...	&	1.93	&	50	&	0.07	&	0.05	&	0.07	&	20.6	&	2008	&	15	\\
OT Ser	&	0.55	&	0.49	&	3.4	&	0.105	&	38.7	&	82.1	&	1.67	&	3.43	&	4.82	&	0.56	&	2008 Feb	&	13\\
V 374 Peg	&	0.28	&	0.28	&	0.5	&	0.007	&	107	&	698	&	2.79	&	39.8	&	81.6	&	0.1	&	2005 Aug	&	16\\
...	&	...	&	...	&	...	&	...	&	94.4	&	616	&	2.46	&	33.3	&	66.9	&	0.12	&	2006 Aug	&	16	\\
WX Uma	&	0.1	&	0.12	&	0.8	&	0.006	&	30.5	&	1090	&	1.07	&	2.06	&	4.91	&	0.2	&	2006	&	15	\\
...	&	...	&	...	&	...	&	...	&	40.4	&	1449	&	1.39	&	3.07	&	7.7	&	0.13	&	2007	&	15	\\
...	&	...	&	...	&	...	&	...	&	40.4	&	1446	&	1.3	&	2.95	&	7.52	&	0.13	&	2008	&	15	\\
...	&	...	&	...	&	...	&	...	&	59.7	&	2137	&	1.61	&	4.69	&	13.2	&	0.08	&	2009	&	15	\\
YZ Cmi	&	0.32	&	0.29	&	2.8	&	0.048	&	115	&	707	&	4.15	&	10.3	&	22	&	0.07	&	2007 Jan	&	14	\\
...	&	...	&	...	&	...	&	...	&	110	&	675	&	4.09	&	9.79	&	20.6	&	0.08	&	2008 Jan	&	14	\\
GJ 176	&	0.49	&	0.47	&	39.3	&	1.059	&	3.62	&	7.26	&	0.15	&	0.01	&	0.01	&	22.6	&	2013 Oct	&	17	\\
GJ 205	&	0.63	&	0.55	&	33.6	&	1.229	&	11.3	&	19.1	&	0.29	&	0.05	&	0.06	&	5.31	&	2013 Oct	&	18	\\
GJ 358	&	0.42	&	0.41	&	25.4	&	0.575	&	43.4	&	133	&	1.11	&	0.34	&	0.58	&	0.44	&	2014 Jan	&	18	\\
GJ 479	&	0.43	&	0.42	&	24	&	0.559	&	11.5	&	33.3	&	0.29	&	0.05	&	0.07	&	3.82	&	2014 Apr	&	18	\\
GJ 674	&	0.35	&	0.4	&	35.2	&	0.66	&	33.5	&	108	&	0.85	&	0.17	&	0.29	&	0.5	&	2014 May	&	17	\\
GJ 846	&	0.6	&	0.54	&	10.7	&	0.369	&	5.31	&	9.39	&	0.14	&	0.05	&	0.06	&	17.1	&	2013 Sept	&	18	\\
...	&	...	&	...	&	...	&	...	&	10.9	&	19.1	&	0.29	&	0.15	&	0.18	&	5.29	&	2014 June	&	18  \\
		\hline
		\end{tabular}
1: Petit et al. (in prep); 2: \citet{Nascimento2014}; 3: \citet{Jeffers2014}; 4: \citet{Petit2008}; 5: \citet{Morgenthaler2011}; 6: \citet{Saikia2015}; 7: \citet{Donati2003Dynamo}; 8: \citet{Folsom2016}; 9: \citet{Fares2009}; 10: \citet{Fares2013}; 11: \citet{Fares2012}; 12: \citet{Fares2010}; 13: \citet{Donati2008}; 14: \citet{Morin2008}; 15: \citet{Morin2010}; 16: \citet{Morin2008V374}; 17: H\'ebrard et al. (in prep); 18: \citet{Hebrard2016}
\end{minipage}
\end{table*}

\subsection{Mass-loss rates}
\label{subsec:MassLoss}
\begin{figure}
	\begin{center}
	\includegraphics[trim = 0.5cm 1.5cm 0.5cm 0cm, width=\columnwidth]{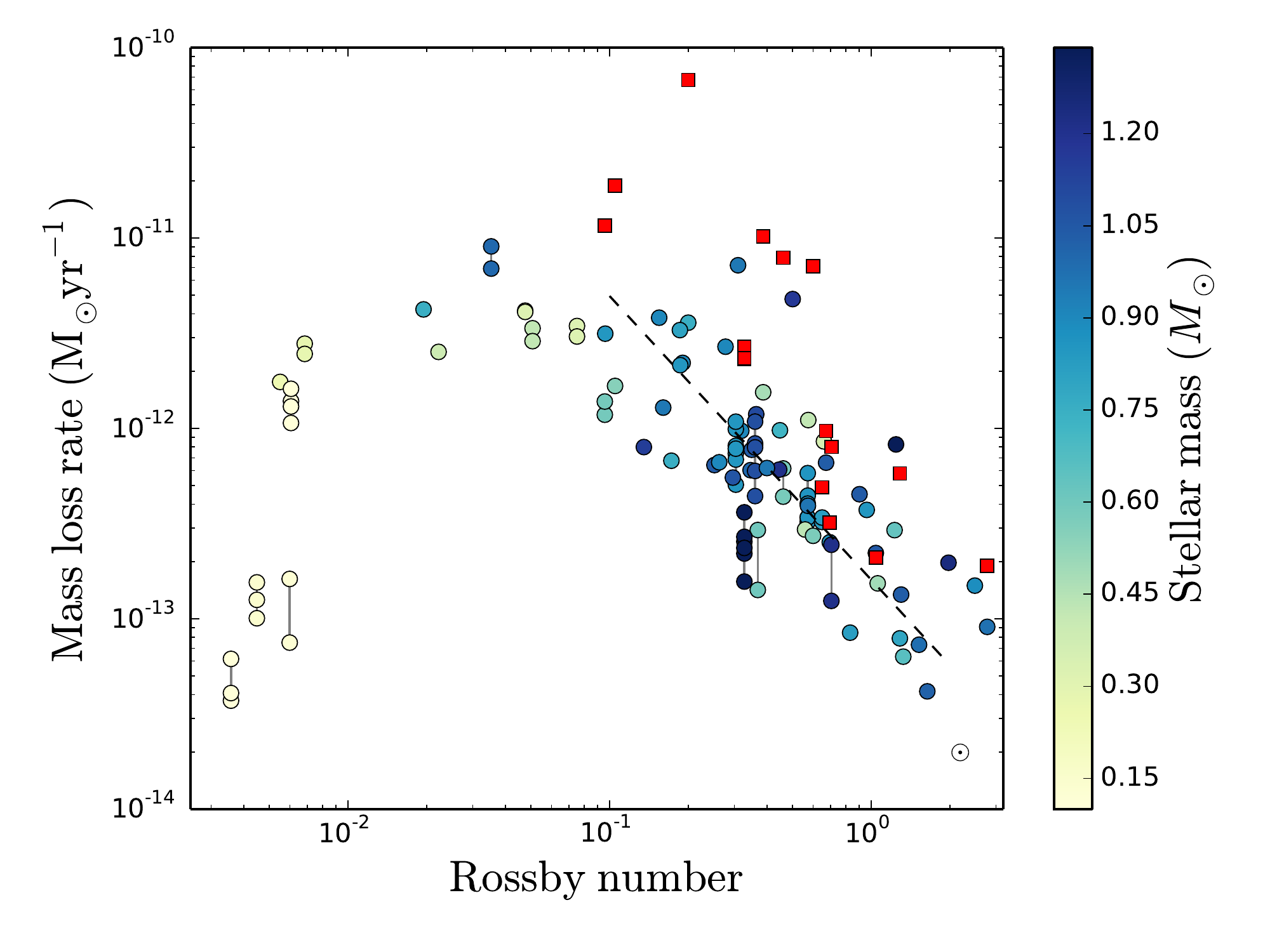}
	\end{center}
	\caption{Predicted mass-loss rate against Rossby number. Symbol colours and meanings are the same as Fig. \ref{fig:OpenFlux}. The fit to the $\rm Ro>0.1$ stars (dashed line) has the form $\dot{M}[M_{\odot}{\rm yr^{-1}}] = (1.60\pm 0.09)\times10^{-13}\ {\rm Ro}^{-1.49\pm0.13}$. Datapoints from 3D MHD simulations (red squares) are not included in the fit.}
	\label{fig:MassLoss}
\end{figure}

In Fig. \ref{fig:MassLoss} we plot the mass-loss rate, as estimated using the method described in section \ref{sec:Model}, against Rossby number for our sample of stars. The mass-loss rates for each star are listed in table \ref{tab:Sample}. As with the open flux, the scatter in the plot is reduced when plotting against Rossby number rather than rotation period (not shown). The fit to the $\rm Ro>0.1$ stars (dashed line) has the form 

\begin{equation}
	\dot{M}{\rm [}M_{\odot}{\rm yr^{-1}]} = (1.60\pm 0.09)\times10^{-13}\ {\rm Ro}^{-1.49\pm0.13}. 
	\label{eq:MDotFit}
\end{equation}
The overall shape of this plot is very similar to that of Fig. \ref{fig:OpenFlux}b ($\Phi_{\rm open}$ vs $\rm Ro$). This is to be expected given the model we have used to estimate the mass-loss rates. The presence of the scaling factor, $f_{\rm mag}$, that accounts for the stronger winds of more active stars, means that the wind density will scale with the surface magnetic field strength. However, the mass-loss rate also has a dependence on the surface area of the star, just as the open flux does, explaining why our predicted mass-loss rates and open fluxes have the same qualitative dependence on Rossby number. Our predicted mass-loss rates are in good agreement with the estimates derived from self-consistent MHD simulations (red squares) at larger Rossby numbers. However, the MHD estimates appear to have a steeper dependence on Rossby number and differ by an order of magnitude or more at smaller Rossby numbers. This is not necesarily surprising given that both models are calibrated using solar values. One way to improve the agreement of the $\dot{M}$ estimates would be to adjust the wind density in the PFSS or MHD model (or both). The wind density is controlled by the $f_{\rm mag}$ parameter in our PFSS model and by the base density of the wind, $n_0$, in the MHD models. Figure \ref{fig:MassLoss} suggests that the ratio of these two model parameters, $n_0/f_{\rm mag}$, is larger for stars with lower Rossby numbers. Better agreement would likely be found if the models were formulated such that $n_0/f_{\rm mag}$ remained constant as a function of Rossby number.

\begin{figure}
	\begin{center}
	\includegraphics[trim = 0.5cm 1.5cm 0.5cm 0cm, width=\columnwidth]{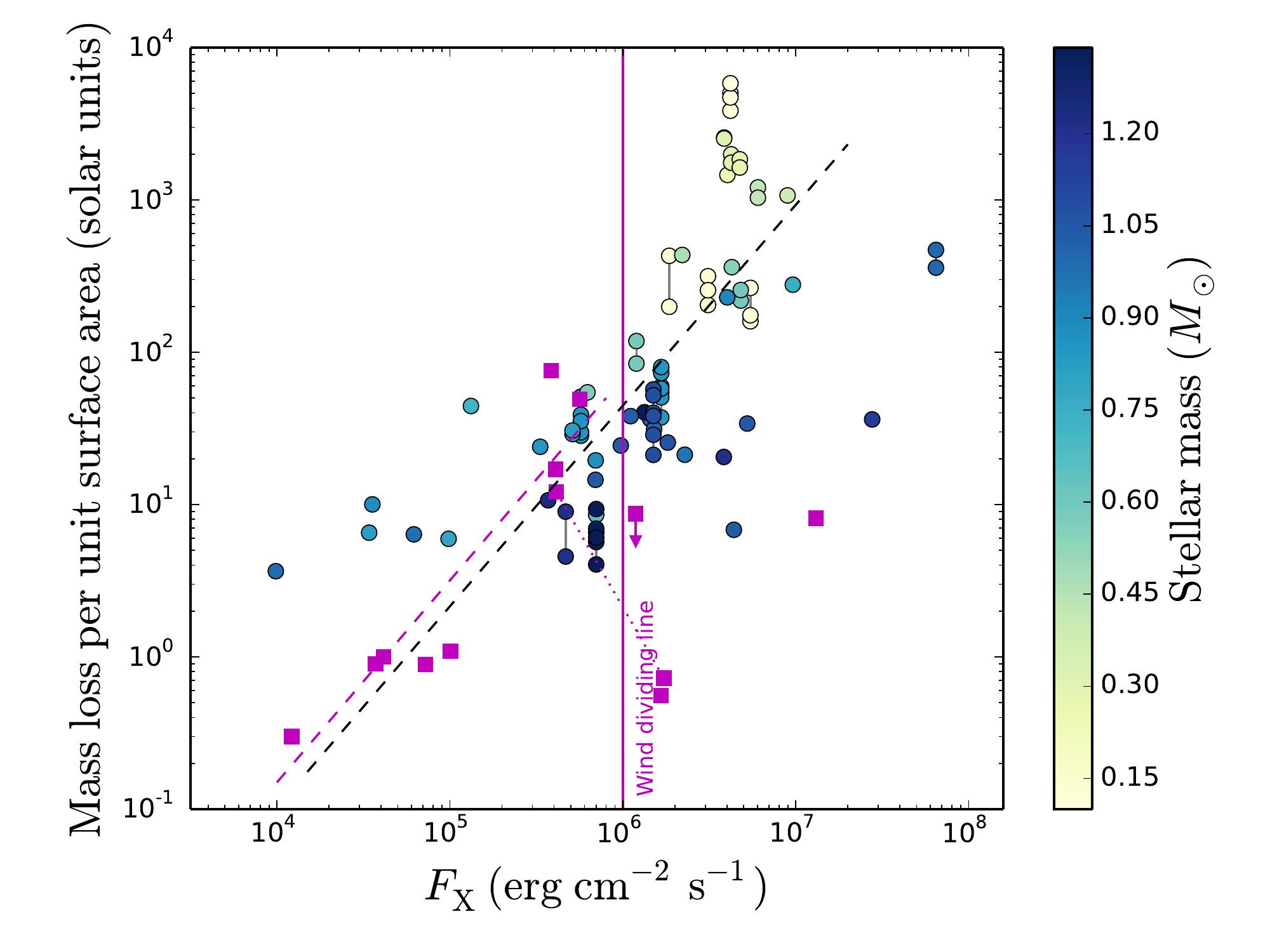}
	\end{center}
	\caption{Mass-loss rate per unit surface area against surface X-ray flux. Symbol colours and meanings are the same as Fig. \ref{fig:OpenFlux}. The black dashed line is a power law fit to the whole sample of the form $\frac{\dot{M}/r_{\star}^2}{\dot{M}_{\odot}/r_{\odot}^2} = (5.51\pm 4.89)\times 10^{-7}\ F_{\rm X}^{1.32\pm 0.15}$. Fitting to the $M_{\star}>0.5M_{\odot}$ stars only, we find a fit of $\frac{\dot{M}/r_{\star}^2}{\dot{M}_{\odot}/r_{\odot}^2} = (5.01\pm 2.53)\times 10^{-4}\ F_{\rm X}^{0.79\pm 0.08}$ (fit line not shown). Overplotted in magenta are the indirect mass-loss rate estimates of \citet[][square symbols]{Wood2014}, their wind dividing line (solid line) and their fit to their stars below the wind dividing line given by, $\dot{M}/r_{\star}^2 \propto F_{\rm X}^{1.34}$ (dashed line).}
	\label{fig:WoodPlot}
\end{figure}

Currently, the best observational constraints on the mass-loss rates of low-mass stars come from indirect estimates based on $\rm Ly\alpha$ observations \citep{Wood2004}. These authors have estimated the mass-loss rates of $\sim10$ stars and find that they have a power law dependence on the surface X-ray flux of $\dot{M}/r_{\star}^2 \propto F_{\rm X}^{1.34}$ \citep{Wood2005}. However, this trend appears to break down for the most active stars. A number of stars with $F_{\rm X} > 10^{6}\ {\rm erg\ cm^{-2}\ s^{-1}}$ seem to have substantially lower mass-loss rates than expected from the stated power law \citep{Wood2014}. These authors suggest that some mechanism, such as a change in magnetic field geometry, inhibits the mass-loss rates of the most active stars and call this divide the `wind dividing line'. In Fig. \ref{fig:WoodPlot}, we reproduce the mass-loss rate vs $F_{\rm X}$ plot of \citet[][their Fig. 4]{Wood2014} with magenta squares as well as their wind dividing line. Additionally, we also plot our mass-loss estimates. A power law fit to our sample with the form $\frac{\dot{M}/r_{\star}^2}{\dot{M}_{\odot}/r_{\odot}^2} = (5.51\pm 4.89)\times 10^{-7}\ F_{\rm X}^{1.32\pm 0.15}$ is shown by the black dashed line. Overall, our mass-loss rates increase with activity, albeit with some scatter. Below the wind dividing line, our mass loss-rates increase with increasing activity which is in qualitative agreement with the results of \citet{Wood2014}. While our mass loss-rates appear to be larger than theirs at the lowest activities, the power law index of our fit, 1.32, is remarkably close to their value of 1.34. However, this may be a coincidence. By eye, the mass-loss rates of the stars in our sample below the wind dividing line clearly having a shallower dependence on $F_{\rm X}$ than the Ly$\alpha$ sample. Additionally, the fit is influenced by the presence of stars with $M_{\star}<0.5M_{\odot}$. These stars are known to have different magnetic field properties, most likely as a result of different internal structures to their higher mass counter parts \citep{Donati2008,Morin2008}. Removing these stars, we obtain a fit of $\frac{\dot{M}/r_{\star}^2}{\dot{M}_{\odot}/r_{\odot}^2} = (5.01\pm 2.53)\times 10^{-4}\ F_{\rm X}^{0.79\pm 0.08}$.

Unlike \citet{Wood2014}, our estimated mass-loss rates continue to increase with increasing activity beyond the wind dividing line. Indeed, we see no substantial change in behaviour in our sample across the $F_{\rm X} = 10^{6}\ {\rm erg\ cm^{-2}\ s^{-1}}$ line. This result is in agreement with previous mass-loss simulations which have also found no change in behaviour over the wind dividing line \citep{See2014,Johnstone2015} as well as the study of \citet{Vidotto2016Wood}. This latter study looked at the large-scale magnetic topologies of all the stars studied by \citet{Wood2014} that had ZDI maps and found that there was no abrupt change in their magnetic field properties over the wind dividing line. Given the small number of stars in the study of \citet{Wood2014}, it remains to be seen whether the idea of a wind dividing line will survive after more stars have had their mass-loss rates estimated with the $\rm Ly\alpha$ technique. On the other hand, if it emerges that stars with $F_{\rm X}>10^6\ {\rm erg\ cm^{-2}\ s^{-1}}$ truly do suffer from reduced mass-loss rates, it would would be an indication that current models of mass-loss are missing important physics. 

\begin{figure*}
	\begin{center}
	\includegraphics[trim = 0cm 1cm 0cm 0cm, width=0.9\textwidth]{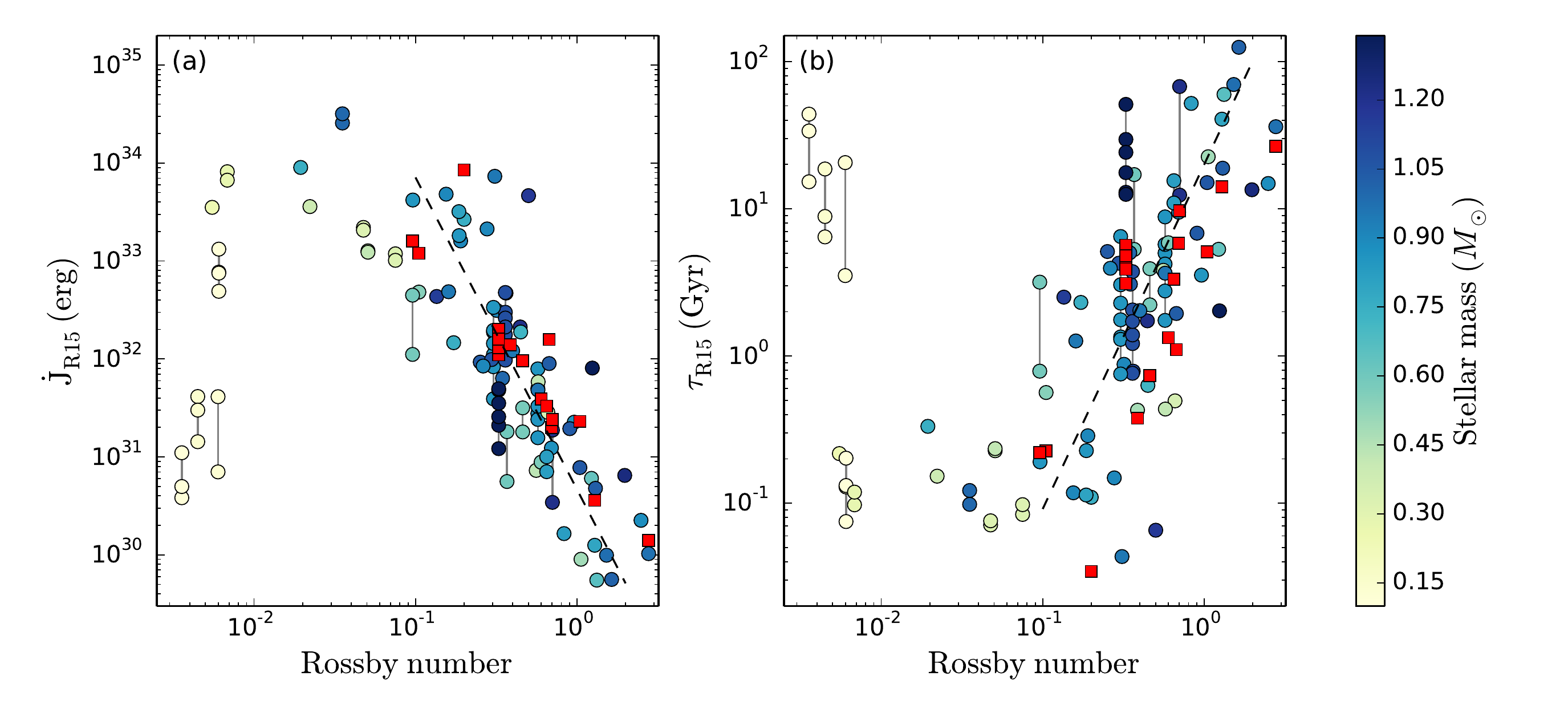}
	\end{center}
	\caption{(a) Angular momentum-loss rates, $\dot{J}_{\rm R15}$ and (b) spin-down time-scale, $\rm \tau_{R15}$, against Rossby number. Symbol colours and meanings are the same as Fig. \ref{fig:OpenFlux}. The fits to the $\rm Ro>0.1$ stars have the form $\dot{J}_{\rm R15}[{\rm erg}] = (4.65\pm 0.47)\times10^{30}\ {\rm Ro}^{-3.19\pm0.25}$ in panel a and $\tau_{\rm R15}[{\rm Gyr}] = (20.0\pm 1.9)\ {\rm Ro}^{2.34\pm 0.22}$ in panel b. Datapoints from 3D MHD simulations (red squares) are not included in the fit.}
	\label{fig:jDotPlot}
\end{figure*}

\begin{figure}
	\begin{center}
	\includegraphics[trim = 0.5cm 1.0cm 0.5cm 0cm, width=\columnwidth]{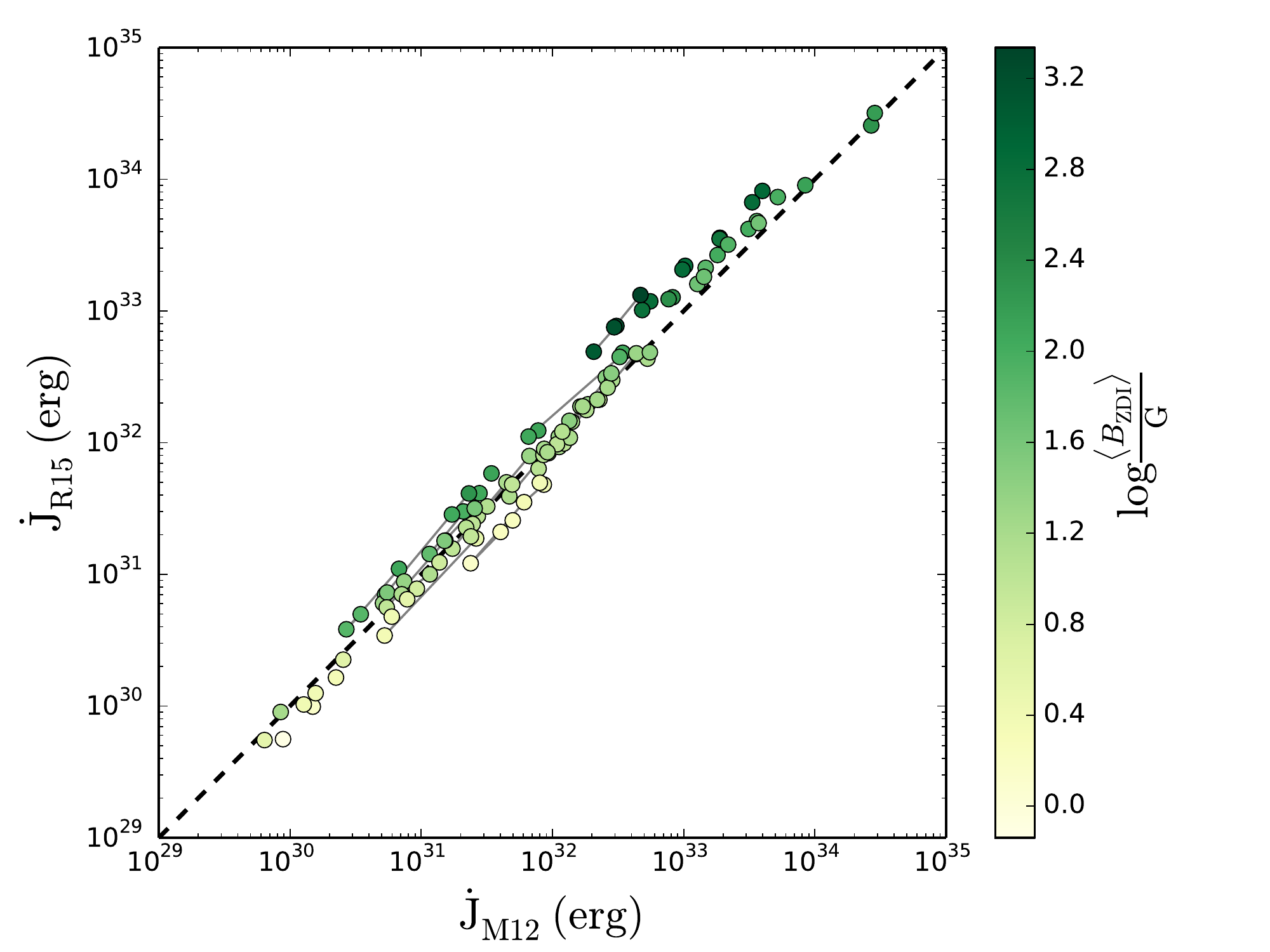}
	\end{center}
	\caption{Comparison of angular momentum-loss rates as estimated by the braking laws of \citet{Matt2012} and \citet{Reville2015}. The dashed line indicates $\dot{J}_{\rm R15}=\dot{J}_{\rm M12}$. Each point is colour coded by the average magnetic field strength at the stellar surface.}
	\label{fig:TorqueComp}
\end{figure}

\subsection{Angular momentum-loss}
\label{subsec:AngMomLoss}
\subsubsection{Trends with Rossby number}
\label{subsubsec:AngMom}
In Fig. \ref{fig:jDotPlot}a, we show the angular momentum-loss rate, as calculated with the R15 braking law (equation (\ref{eq:Reville15})), against Rossby number. The angular momentum-loss rate shows a similar qualitative behaviour to the open flux (Fig. \ref{fig:OpenFlux}b) and mass-loss rate (Fig. \ref{fig:MassLoss}); increasing $\dot{J}_{\rm R15}$ with decreasing Rossby number and a large spread in $\dot{J}_{\rm R15}$ below $\rm Ro\sim 0.01$. This is unsurprising given the dependence of the angular momentum-loss rate on the open flux and mass-loss rate. The fit to the $\rm Ro>0.1$ stars (dashed line) has the form 

\begin{equation}
	\dot{J}_{\rm R15}[{\rm erg}] = (4.65\pm 0.47)\times10^{30}\ {\rm Ro}^{-3.19\pm0.25}. 
	\label{eq:JDotFit}
\end{equation}
In Fig. \ref{fig:jDotPlot}b, we show the instantaneous spin-down time-scale, $\tau_{\rm R15}=J/\dot{J}_{\rm R15}$ for our sample of stars. Values for both $\dot{J}_{\rm R15}$ and $\tau_{\rm R15}$ are listed in table \ref{tab:Sample}. In order to calculate the angular momentum, $J=I\Omega_{\star}$, we require the moments of inertia, $I$, for each star. These were estimated using the evolutionary models of \citet{Baraffe2015} at an age of 500 Myr. This should be representative of the moments of inertia for these stars since they do not evolve significantly over their main sequence lifetime. The fit to the $\rm Ro>0.1$ stars (dashed line) has the form $\tau_{\rm R15} = (20.0\pm 1.9)\ {\rm Ro}^{2.34\pm 0.22}$. The angular momentum-loss rates and spin-down time-scales\footnote{Our method of estimating the stellar angular momentum, $J$, differs from the method used by the MHD simulations. Therefore, we have recalculated the spin-down time-scales for the MHD simulations using our angular momentum values to provide a consistent comparison to our model results.} estimated from MHD simulations (red squares) agree well with our results. This is in spite of the fact that the PFSS and MHD estimates for the mass-loss rates used to calculate the angular momentum-loss rate show larger disagreements (section \ref{subsec:MassLoss}). The reason for this is that the angular momentum-loss rate (equation (\ref{eq:Reville15})) has a much stronger dependence on the open flux, $\dot{J}_{\rm R15}\propto \Phi^{4m_2}_{\rm open}\propto \Phi^{1.24}_{\rm open}$, compared to the mass loss-rate, $\dot{J}_{\rm R15}\propto \dot{M}^{1-2m_2}\propto \dot{M}^{0.38}$.

The angular momentum-loss rates of stars with $\rm Ro>0.01$ decreases as a function of Rossby number. Correspondingly, the spin-down time-scale increases with Rossby number. This is a simple consequence of stellar activity declining as stars spin down over their lifetimes. However, the spin-down behaviour of stars with $\rm Ro<0.01$ is much more intriguing. This regime consists of all the $M_{\star}<0.2\solarmass$ stars whose magnetic fields exist in one of two distinct states \citep{Morin2010}, as well as a number of slightly higher mass stars (all $<0.28 \solarmass$). The two distinct set of magnetic characteristics has resulted in two groups of stars. Those with strong dipolar fields (V374 Peg, EQ Peg B \& WX UMa) have large angular momentum-loss rates and lie roughly at the tail end of the sequence of $\rm Ro > 0.01$ stars with spin-down time-scales of $\tau_{\rm R15}\sim 100-200$ Myr. However, the stars with weak and more complex field structures (DX Cnc, GJ 1156 \& GJ 1245B) have angular momentum-loss rates that are two orders of magnitude weaker compared to their strong field counterparts, despite having comparable Rossby numbers. These stars have much longer spin-down scales on the order of $\tau_{\rm R15}\sim 3-40$ Gyr. It is perhaps odd that the stars in both the strong and weak field states have similar rotation periods despite the large difference in the spin-down time-scales. A coherent theory of angular momentum evolution of these types of stars will need to explain how this is possible.

\subsubsection{Comparing braking laws}
\label{subsubsec:BreakingLaws}
At this stage, it will be instructive to compare the M12 and R15 braking laws and their ability to predict the angular momentum-loss rate of a star. The R15 braking law is formulated to account for the effects of complex magnetic field geometries whereas the M12 braking law was constructed by considering only dipolar geometries. However, the R15 braking law is expressed in terms of the open flux while the M12 braking law is expressed in terms of the surface field strength of the dipolar component. While better capturing the effects of complex field geometries, this means that the R15 law is also, in practice, more difficult to use since the open flux of a star is a more difficult quantity to obtain or estimate. In particular, when using the PFSS model, accurately estimating the open flux of a star will require an accurate estimate of the source surface radius which is a free parameter of the model.

In section \ref{subsec:CoronalField}, we showed that the open flux is dominated by the dipole component of the field which suggests that the M12 braking law is sufficient to provide an accurate estimate of the angular momentum-loss rate of a star. In Fig. \ref{fig:TorqueComp}, we compare the angular momentum-loss rates as calculated by the formulations of M12 and R15, the values of which are listed in table \ref{tab:Sample}. When calculating $\dot{J}_{\rm R15}$ and $\dot{J}_{\rm M12}$, the open flux, as shown in Fig. \ref{fig:OpenFlux}, and the average surface field strengths of the dipolar component, as shown in Fig. \ref{fig:DipoleVsOpenFlux}, are used respectively. The two formulations agree very well with most stars agreeing to within a factor of $\sim 1.5$. Even the stars with the largest discrepancies still agree to within a factor of $\sim 3$. 

When interpreting Fig. \ref{fig:TorqueComp}, there are a number of factors of which we must be mindful. Firstly, our result that the open flux is predominantly determined by the dipolar component of the field is based on the assumed source surface radii of $r_{\rm ss}=3.41r_{\star}$ for all our stars. However, the source surface radius is a free parameter within our model. If the source surface radius is actually smaller than our chosen value, higher order multipoles may have a non-negligible effect on the open flux making the M12 formulation less accurate. Secondly, the R15 formulation is also dependent on our choice of source surface radius. For a given input ZDI map, the open flux decreases for larger choices of the source surface radius and vice versa. In reality, one would expect the source surface radius to vary as a function of the fundamental parameters of a star \citep{Reville2015b}. 

It is interesting to note that there is a systematic trend with the average surface magnetic field strength, $\langle B_{\rm ZDI}\rangle$, in Fig. \ref{fig:TorqueComp}. The $\dot{J}_{\rm R15}>\dot{J}_{\rm M12}$ stars have the strongest surface magnetic fields, and, conversely, $\dot{J}_{\rm R15}<\dot{J}_{\rm M12}$ stars have the weakest. It is not unreasonable to suggest that stars with stronger magnetic fields may have larger source surface radii. Increasing the source surface radii of the $\dot{J}_{\rm R15}>\dot{J}_{\rm M12}$ (strong field) stars would reduce their open flux and bring their $\dot{J}_{\rm R15}$ estimates closer to their $\dot{J}_{\rm M12}$ estimates. Conversely, reducing the source surface radii of the $\dot{J}_{\rm R15}<\dot{J}_{\rm M12}$ (weak field) stars would increase their open flux and also bring their $\dot{J}_{\rm R15}$ estimates closer to their $\dot{J}_{\rm M12}$ estimates. Of course, we should remember that our aim is not necessarily to make the $\dot{J}_{\rm R15}$ estimates match the $\dot{J}_{\rm M12}$ estimates since, as we have already discussed, the $\dot{J}_{\rm M12}$ formulation will become less accurate for stars where higher order field mores are not negligible.

It is clear that the source surface radius is an important parameter to set properly. Fig. \ref{fig:TorqueComp} shows that the M12 and R15 formulations are in reasonable agreement. It is likely that both of these formulations, as implemented in this work, can provide a reasonably reliable estimate of the angular-momentum-loss rate of a star. This is especially true when considering that the discrepancies between the M12 and R15 formulations are much smaller than the scatter in the physical properties input into these formulations, e.g. open flux. However, the issues of higher order field modes and our  assumption of a constant source surface radius means that neither formulation, as implemented in this work, are perfect estimates. Further investigation will be required to determine when the M12 formulation is valid to use (since it is the easier formulation to use) and what source surface radius to use when the R15 formulation is used.

\subsection{Comparison to 3D MHD simulations}
\label{subsec:Comprison}
MHD simulations have an advantage over the model used in this work because they include more self-consistent physics at the cost of being much more computationally expensive. A computationally cheaper model, such as the one we have presented here, is therefore required to model this many stars within a reasonable time frame. Both types of model suffer from the fact that wind densities are difficult to constrain observationally. With these points in mind, it is important to study the differences in the models in order to understand how the choice of model affects the end results.

The MHD simulations that we compare to make use of the BATS-R-US numerical code \citep{Powell1999}. This is a 3D code that solves the ideal MHD equations on an adaptive mesh grid in Cartesian coordinates. The simulations require an initial magnetic field and stellar wind configuration. The wind is initialised using a thermally driven Parker wind \citep{Parker1958} while the 3D magnetic field structure is initialised using a ZDI map and the PFSS model. The wind and field are then allowed to interact self-consistently until the simulation relaxes to a steady state solution. As discussed in section \ref{subsec:MassLoss}, the base wind density are a free parameter within these simulations.

We have demonstrated that our results broadly agree with those of the MHD simulations. It is clear that the open flux (Fig. \ref{fig:OpenFlux}), angular momentum-loss rates (Fig. \ref{fig:jDotPlot}a) and instantaneous spin-down time-scales (Fig. \ref{fig:jDotPlot}b) calculated from MHD simulations lie within the scatter of the results from our own models. While the mass-loss estimates agree reasonably well for low activity stars, the agreement becomes worse at lower Rossby numbers (Fig. \ref{fig:MassLoss}). This is likely due to the different choices of wind densities in the models (see section \ref{subsec:MassLoss}). On the whole, our model appears to produce reasonable values and trends for these parameters.

As well as comparing overall trends, we can make a direct comparison of the models for each of the stars that have been modeled with an MHD code. In table \ref{tab:MHDComp}, we list estimates for the open flux, mass-loss rate, angular momentum-loss rate and instantaneous spin-down time-scale from both models, as well as the ratio of the values obtained under both models. The open fluxes show a very good agreement - within a factor of $\sim 2$ for the large majority of the stars. The mass loss-rates on the other hand match less well. In some cases, the mass loss-rates obtained from our PFSS model and the MHD models differ by an order of magnitude or more. In particular, the discrepancy is larger for M dwarf stars. As discussed in section \ref{subsec:MassLoss}, a better agreement between the models could be found by adjusting the wind densities in each of the models. The angular momentum-loss rates, $\dot{J}$, and spin-down time-scales, $\tau$, both show a reasonable agreement. These values agree to within a factor of $\sim 4$ for the majority of the stars.  As with the mass loss-rates, the angular momentum-loss rates estimated from our PFSS model are lower than the values obtained from the MHD models. Given the dependence of $\dot{J}$ on $\dot{M}$ (see equation (\ref{eq:Reville15})), reducing the discrepancy in $\dot{M}$ between the models would also decrease the discrepancy in $\dot{J}$.

\begin{landscape}
 \begin{table}
	\caption{Comparison of the results from the our model (denoted as ``PFSS") with the results from 3D MHD models (denoted as ``MHD"). For each star, we list the open flux, $\Phi_{\rm open}$, mass-loss rate, $\dot{M}$, angular momentum-loss rate, $\dot{J}$, and instantaneous spin-down time-scale, $\tau$, from our model and the MHD models. Additionally, we list the ratio of each of these quantities from the two models given by dividing the MHD model value by the PFSS model value. References indicate the original article in which the MHD models were published. It should be noted that the MHD spin-down time-scales, $\tau_{\rm MHD}$, have been recalculated and are not the same as the values published in the original articles (see discussion in section \ref{subsec:Comprison}).}
	\label{tab:MHDComp}
	\begin{tabular}{lccccccccccccc}
		\hline
		Star & $\Phi_{\rm open}^{\rm PFSS}$ & $\Phi_{\rm open}^{\rm MHD}$ & ratio ($\Phi_{\rm open}$) & $\dot{M}_{\rm PFSS}$ & $\dot{M}_{\rm MHD}$ & ratio ($\dot{M}$) & $\dot{J}_{\rm R15}^{\rm PFSS}$ & $\dot{J}^{\rm MHD}$ & ratio ($\dot{J}$) & $\tau_{\rm R15}^{\rm PFSS}$ & $\tau^{\rm MHD}$ & ratio ($\tau$) & Ref.\\
		ID & ($10^{22}$Mx) & ($10^{22}$Mx) & & ($10^{-12}M_{\odot}yr^{-1}$) & ($10^{-12}M_{\odot}yr^{-1}$) & & ($10^{32}$erg) & ($10^{32}$erg) & & (Gyr) & (Gyr) & & \\
		\hline
HD 189733 (2008 July)	&	6.01	&	11.14	&	1.85	&	0.32	&	0.49	&	1.53	&	0.1	&	0.33	&	3.30	&	10.94	&	3.32	&	0.30	&	1	\\
GJ 49	&	10.1	&	14.56	&	1.44	&	0.27	&	7.1	&	26.30	&	0.088	&	0.39	&	4.43	&	5.88	&	1.33	&	0.23	&	2	\\
CE Boo	&	45.9	&	50.6	&	1.10	&	1.55	&	10.2	&	6.58	&	1.24	&	1.4	&	1.13	&	0.43	&	0.38	&	0.89	&	2	\\
DS Leo (2007 Dec)	&	12.2	&	17.55	&	1.44	&	0.44	&	7.88	&	17.91	&	0.18	&	0.96	&	5.33	&	3.92	&	0.73	&	0.19	&	2	\\
GJ 182	&	104	&	150	&	1.44	&	3.59	&	67.6	&	18.83	&	26.6	&	85	&	3.20	&	0.11	&	0.034	&	0.31	&	2	\\
OT Ser 	&	38.7	&	62.4	&	1.61	&	1.67	&	18.8	&	11.26	&	4.8	&	12	&	2.50	&	0.56	&	0.23	&	0.41	&	2	\\
DT Vir (2007 Jan)	&	33.8	&	28.2	&	0.83	&	1.18	&	11.6	&	9.83	&	4.48	&	16	&	3.57	&	0.79	&	0.22	&	0.28	&	2	\\
HD 46375	&	3.51	&	4.42	&	1.26	&	0.09	&	0.19	&	2.11	&	0.01	&	0.014	&	1.40	&	36.11	&	26.5	&	0.73	&	3	\\
HD 73256	&	5.59	&	5.46	&	0.98	&	0.22	&	0.21	&	0.95	&	0.078	&	0.23	&	2.95	&	15.06	&	5.09	&	0.34	&	3	\\
HD 102195	&	7.19	&	8.61	&	1.20	&	0.25	&	0.32	&	1.28	&	0.12	&	0.2	&	1.67	&	9.47	&	5.84	&	0.62	&	3	\\
HD 130322	&	2.87	&	5.11	&	1.78	&	0.08	&	0.58	&	7.25	&	0.013	&	0.036	&	2.77	&	40.62	&	14.1	&	0.35	&	3	\\
HD 179949 (2007 June)	&	1.73	&	4.42	&	2.55	&	0.12	&	0.8	&	6.67	&	0.034	&	0.24	&	7.06	&	67.66	&	9.68	&	0.14	&	3	\\
HD 20630	&	19.56	&	22.5	&	1.15	&	0.66	&	0.97	&	1.47	&	0.9	&	1.58	&	1.76	&	1.94	&	1.1	&	0.57	&	4	\\
$\tau$ Boo (2009 May)	&	4.45	&	4.5	&	1.01	&	0.24	&	2.34	&	9.75	&	0.35	&	1.3	&	3.71	&	17.60	&	4.78	&	0.27	&	5	\\
$\tau$ Boo (2010 Jan)	&	5	&	7.6	&	1.52	&	0.36	&	2.31	&	6.42	&	0.48	&	2	&	4.17	&	12.94	&	3.11	&	0.24	&	5	\\
$\tau$ Boo (2011 Jan)	&	5.72	&	5.6	&	0.98	&	0.25	&	2.34	&	9.36	&	0.49	&	1.6	&	3.27	&	12.52	&	3.89	&	0.31	&	5	\\
	\hline
	\end{tabular}
    	\hspace{2cm} 1: \citet{Llama2013}; 2: \citet{Vidotto2014Torque}; 3: \citet{Vidotto2015}; 4: \citet{Nascimento2016}; 5: \citet{Nicholson2016}
 \end{table}
\end{landscape}

\noindent Overall, our computationally efficient PFSS model provides a reasonable match to the MHD models although there is still room for improvement.

\section{Discussion and conclusions}
\label{sec:Conlusions}
In recent years, large strides have been made in understanding the factors that affect the rotational evolution of main sequence stars. In particular, the dependence of the angular momentum-loss rate on the open flux of a star has been quantified through the use of MHD simulations \citep{Vidotto2014Torque,Reville2015}. Formulating a braking law in terms of the open flux is an improvement over previous implementations since it accounts for the complex magnetic field geometries that stars are known to have. In principle, such a braking law can be used to compute the rotation period evolution of a star over its main sequence lifetime. However, there have been little to no systematic studies investigating how the open flux varies as a function of fundamental stellar parameters.

In this work, we use a sample of stars that have had their large-scale surface magnetic fields mapped using the ZDI technique to study various parameters related to stellar spin-down. This is the largest sample of ZDI maps assembled in one study to date. Using the potential field source-surface  method, we analyse how the open flux of these stars, as well as their mass loss-rates, angular momentum-loss rates and instantaneous spin-down time-scales vary as a function of fundamental stellar parameters. The choice to use a potential field source-surface model was driven by the requirement of a computationally efficient method with which to model our large sample. Indeed, attempting to model all our stars with a multi-dimensional MHD model would have taken a prohibitively long amount of time. Our results indicate that the potential field source-surface model provides results that are in reasonable agreement with MHD models.

We find that the open flux for the majority of our stars is predominantly determined by the dipole component of the magnetic field for our choice of the source surface radii ($r_{\rm ss}=3.41r_{\star}$). Similarly to previous studies of other proxies of magnetic activity \citep[e.g.][]{Wright2011}, the open flux, mass loss-rates and angular momentum-loss rates all show less scatter when plotted against Rossby number rather than rotation period. In the unsaturated regime ($\rm Ro \gtrsim 0.1$), each of these parameters increase with decreasing Rossby number. At $\rm Ro \lesssim 0.1$, magnetic proxies typically saturate. We also see evidence of saturation in the open flux, mass loss-rates and angular momentum-loss rates although it is not as clear as it is for other magnetic proxies because of a dearth of stars in our sample in this regime. When comparing to mass-loss rates estimated from the Ly$\alpha$ technique, we do not predict a drop in mass-loss rates above the wind dividing line as \citet{Wood2014} suggest. Further work, will be required to determine the reason for this difference in mass-loss rate behaviour at the highest activities. Clarifying how mass-loss rates evolve over time, especially early on in the lifetime of a star, will be crucial to understanding whether potentially habitable planets can retain their atmospheres \citep[e.g.][]{Ribas2016}. 

At the smallest Rossby numbers ($\rm Ro \lesssim 0.01$), the behaviour of the open flux, mass- and angular momentum-loss rates changes abruptly. Due to the presence of both strong dipolar magnetic fields and weak multipolar fields in the lowest mass stars \citep{Morin2010}, the open flux, mass loss-rates, angular momentum loss-rates and spin-down time-scales of these stars are spread out over many orders of magnitude over a very narrow range of Rossby numbers. Models of angular momentum evolution \citep[e.g.][]{Reiners2012,Matt2015} typically do not account for both the strong and weak field states in the lowest mass M dwarfs. The model of \citet{Reiners2012} predicts long spin-down time-scales for the lowest mass stars due to their small stellar radii. Our model also predicts long spin-down time-scales for the low mass M dwarfs but only for the weak multipolar stars. Unlike the model of \citet{Reiners2012}, we simultaneously predict a population of low mass M dwarfs with much shorter spin-down time-scales, i.e. those with strong dipolar fields.

A number of suggestions have been proposed to explain the two magnetic field states observed for the lowest mass M dwarfs. The first is that a parameter other than rotation and mass, such as age, may have an effect. Indeed, \citet{Morin2010} discuss the possibility these stars switch from a weak multipolar state to a strong dipolar state at some point in their lifetimes. This suggestion is supported by the fact that the weak field stars in their sample belong to a young kinematic population while the strong field stars belong to older kinematic populations. Another suggestion is that these stars switch repeatedly between the two states as part of some form of magnetic cycle on the time-scale of decades \citep{Morin2010,Kitchatinov2014}. Finally, it may be the case that these stars occupy a bistable region of parameter space such that two different stable field configurations are possible for a very similar set of stellar parameters \citep{Morin2011,Gastine2013}. Under this scenario, the dynamos of these stars would be capable of generating either strong or weak surface magnetic fields but would not be expected to switch between them. 

The spread in field strengths, and hence angular momentum-loss rates, at Rossby numbers less than 0.01 has implications for the rotation period evolution of these stars. Such a large spread in spin-down times may go some way to explaining the spread in rotation periods observed in low-mass M dwarfs \citep[e.g.][]{Irwin2011,Newton2016,Douglas2016}. Additionally, each of the scenarios discussed should produce different rotation period distributions at late ages. For instance, under the bistability scenario, one might expect to observe a population of fast rotators and a population of slow rotators. On the other hand, the effects of a star switching between strong and weak field states should average out over evolutionary time-scales, possibly resulting in these stars converging to a narrow range of rotation periods at late ages. Finally, as part of the MEarth project, \citet{Newton2016} measured the rotation periods of several hundred mid M dwarfs and found the distribution to be consistent with a scenario of rapid rotation for several Gyr followed by rapid spin-down. Such a scenario would be compatible with the idea that these stars switch from a weak field to strong field state at some point within their lifetimes. It is also curious that a significant fraction of the $\rm Ro<0.01$ stars are in the strong dipole field state considering how much shorter their spin-down time-scales are when compared to the weak field counterparts. The fact that the exact ages of the strong/weak field stars in our sample are unknown prevent us from conducting a detailed study of them in the context of rotation period data from clusters of known ages.

While significant progress has been made in the field of rotational period evolution of main sequence stars, there still remain open questions. For instance, what is the long term magnetic evolution of $\rm Ro<0.01$ M dwarfs? Further rotation period measurements of fully concective M dwarfs and further ZDI reconstructions of M dwarfs in this regime will be required to answer this question. In this context, SPIRou, a nIR spectropolarimeter in construction for the Canada-France-Hawaii telescope \citep[][first light expected in 2017]{Moutou2015}, should greatly improve our understanding, especially for fully convective M dwarfs with long rotation periods. Additionally, under what circumstances is the dipole component of the magnetic field enough to accurately determine the open flux of a star? Answering these and other questions will be important in developing a truly holistic understanding of the coherent rotational period evolution seen in open clusters.

\section*{Acknowledgements}
The authors thank an anonymous referee for their constructive comments that helped improve this work. VS acknowledges the support of a Science \& Technology Facilities Council (STFC) postdoctoral fellowship. SBS and SVJ acknowledge research funding by the Deutsche Forchungsgemeinschaft (DFG) under grant SFB, project A16. This study was supported by the grant ANR 2011 Blanc SIMI5-6 020 01 ``Toupies: Towards understanding the spin evolution of stars" (\url{http://ipag.osug.fr/Anr_Toupies/}). This work is based on observations obtained with ESPaDOnS at the CFHT and with NARVAL at the TBL. CFHT/ESPaDOnS are operated by the National Research Council of Canada, the Institut National des Sciences de l'Univers of the Centre National de la Recherche Scientifique (INSU/CNRS) of France and the University of Hawaii, while TBL/NARVAL are operated by INSU/CNRS. We thank the CFHT and TBL staff for their help during the observations.

\bibliographystyle{mnras}
\bibliography{OpenfluxPaper}{}

\end{document}